\documentclass[manuscript,showpacs,preprintnumbers,amsmath]{revtex4}

\usepackage[dvips]{graphicx}
\usepackage{epsf}
\usepackage{amssymb}
\usepackage{psfrag}

\usepackage{graphicx,float,psfrag,subfigure}
\usepackage{dcolumn}
\usepackage{bm}

\begin{document}

\title{Dynamical structures in binary media of potassium-driven neurons}

\author{D. E. Postnov$^1$} \author{F. M\"uller$^2$} \author{R. B. Schuppner$^2$} \author{L. Schimansky-Geier$^2$}%

\affiliation{$^1$Department of Physics, Saratov State University, Astrakhanskaya ul. 83, Saratov 410012,  Russia.\\
  $^2$Institute of Physics, Humboldt-University at Berlin, Newtonstr.
  15, D-12489 Berlin, FR Germany}


\begin{abstract}
  According to the conventional approach neural ensembles are modeled
  with fixed ionic concentrations in the extracellular environment.
  However, in some cases the extracellular concentration of potassium
  ions can not be regarded as constant. Such cases represent specific
  chemical pathway for neurons to interact and can influence strongly
  the behavior of single neurons and of large ensembles. The released
  chemical agent diffuses in the external medium and lowers thresholds
  of individual excitable units. We address this problem by studying
  simplified excitable units given by a modified FitzHugh-Nagumo
  dynamics.  In our model the neurons interact only chemically via the
  released and diffusing potassium in the surrounding non-active
  medium and are permanently affected by noise. First we study the
  dynamics of a single excitable unit embedded in the extracellular
  matter. That leads to a number of noise-induced effects, like
  self-modulation of firing rate in an individual neuron. After the
  consideration of two coupled neurons we consider the spatially
  extended situation. By holding parameters of the neuron fixed
  various patterns appear ranging from spirals and traveling waves to
  oscillons and inverted structures depending on the parameters of the
  medium.
\end{abstract}

\maketitle

\noindent

\newpage
\section{Introduction}
According to a conventional view on neuronal dynamics, the electrical
activity of a cell is represented by depolarization of its membrane
potential.  It is well described by the famous Hodgkin-Huxley model
\cite{HH} describing the dynamics of the different ion channels and
the gating ions.  But also simplified model systems using less number 
of variables and control parameters \cite{keener}
are able to specify the many aspects of neuronal dynamics. Among
them, the FitzHugh-Nagumo model \cite{FHN} with a few control
parameters plays the role of popular paradigm for many excitable
systems. Due to its simplicity it helps a lot that ideas, which have
been developed in neuronal dynamics, can be transferred to related
problems of other nonlinear dynamics \cite{review_lindner}.

But an inevitable assumption and simplification has been made for
these prototypical neuron models in terms of ionic currents. For
example, it was assumed that in spite of transmembrane currents, both
extracellular and intra-cellular ionic concentrations remain unchanged
during depolarizations.  Such a simplification is natural and
acceptable if one considers individual neurons or segments of an
excitable medium during a sufficiently short course of firing.

However, in other cases it will be not realistic. For example, there
is experimental evidence that extracellular concentration of
potassium ions can vary significantly during the course of intensive
neuronal firings \cite{sykova_83, Dahlem_03}.  The detailed modeling
of what happens during the course of ischemia shows the strong
increase of extracellular potassium concentration up to 80 mM
\cite{Yi_2003,yan_96,hansen_78,barreto_1,barreto_2}.  The glial cells, surrounding and
supporting neurons, activate the potassium pumping when its
concentration rise considerably (more than twice in medical leech)
\cite{Deitmer}.  This excessive elevation of potassium concentration
is considered to be an important element of mechanism of epileptic
seizure development \cite{Bazhenov1,Bazhenov2}.
As the relevant computational studies, early modeling attempts were
focused mostly on mechanisms of extracellular potassium clearance and
showed that pathways different from diffusion must be involved in this
process \cite{Vern_1977,Gardner-Medwin_1983,Odette_1988,Dietzel_1989}.
More recent models addressed the detailed neuronal morphology
\cite{Kager_2000,Kager_2002} or the role of specific ion
channels in formation of self-sustained bursting behavior
\cite{Bazhenov1,Frohlich_2006}.  It was also shown that the
interplay between ion concentrations and neural activity can lead to
self-sustained pathological neural activation even in the case of an
isolated cell.  The wider list of modeling issues on topic was
recently reviewed in \cite{Frohlich_2008}.

While the effects of variable ionic concentration are embedded in the
quantitative high-dimensional models, it is difficult to distinguish
them from other aspects of system behavior.  At the same time, the
set of frequently used simplified models does not cover the problem,
just not having the appropriate control parameters.

One of the authors addressed this problem in recent works
\cite{potass_1, potass_2}, where the effects of potassium mediated
coupling were investigated
using the Hodgkin-Huxley type model of leech neurons.  It was shown,
that such rather simplified, but still quantitative model reproduces
the main features of small ensembles of potassium-driven neurons.

However, to study the behavior of large networks by means of a
quantitative model, one need to reduce the number of control
parameters that are difficult to estimate or just unknown.  The
alternative way is to develop a simplified non-dimensional model that
can capture at a qualitative level the specific features of
potassium-coupled neurons and allows one to build large networks
using reasonable set of control parameters.

In the present work we derive such a model in the form of an extended
FitzHugh-Nagumo (FHN) system \cite{FHN} with an additional equation
describing the dynamics of extracellular potassium.  Since our model
inherits the key features of a FHN neuron, it is physically
transparent and tractable and thus provides the better chance to learn
more about nonlinear mechanisms governing the formation of
spatio-temporal patterns in large networks. Furthermore,
compared to leaky integrate and fire models (LIF) the FHN includes
the reset mechanism of the neuron. Some dynamical behavior 
which we will observe later on in the article refers
to the excited state of the FHN which is not provided by the LIF-model
\cite{coombes_03,coombes_06}.

Applied to interacting neurons we assume that the interaction of
neurons is restricted to the single chemical pathway. Coupling takes
place indirectly due to the potassium concentration outside the cells,
only.  Although there is no explicit distance defined in the extended
model, the neurons are strictly separated and a direct contact of the
action potentials is excluded. Therefore the transport of activity
through the heterogeneous medium of neurons and exterior is slower
than in a homogeneous excitable medium.

The patterns appearing in the two-dimensional system show
phenomena, which are reminiscent of chemical experiments in which
comparable heterogeneous situations like two-layer systems or chemical
oscillators moving in a diffusive environment have been studied
\cite{Epstein, Showalter}. Also some of the presented structures
relate to studies on the $Ca^{2+}$ release across the endoplasmathic
recituculum \cite{Falcke,Jung}, where clusters with a finite number
of ion channels on the recituculum take over the role of the noisy
excitable neurons and $Ca^{2+}$ diffuses freely in the cytoplasm.

In our work we attempt to classify the observed spatio-temporal
patterns according to the relation between the key control parameters
of the extracellular medium.  The most noticeable behavior are
randomly-walking spots, long living meandering excitations, anti-phase
firing patterns and inverted spirals and waves.

\section{Minimal model for a potassium-driven neuron}
\subsection{Background}
\begin{figure}[t] \centering
  \includegraphics[width=\textwidth]{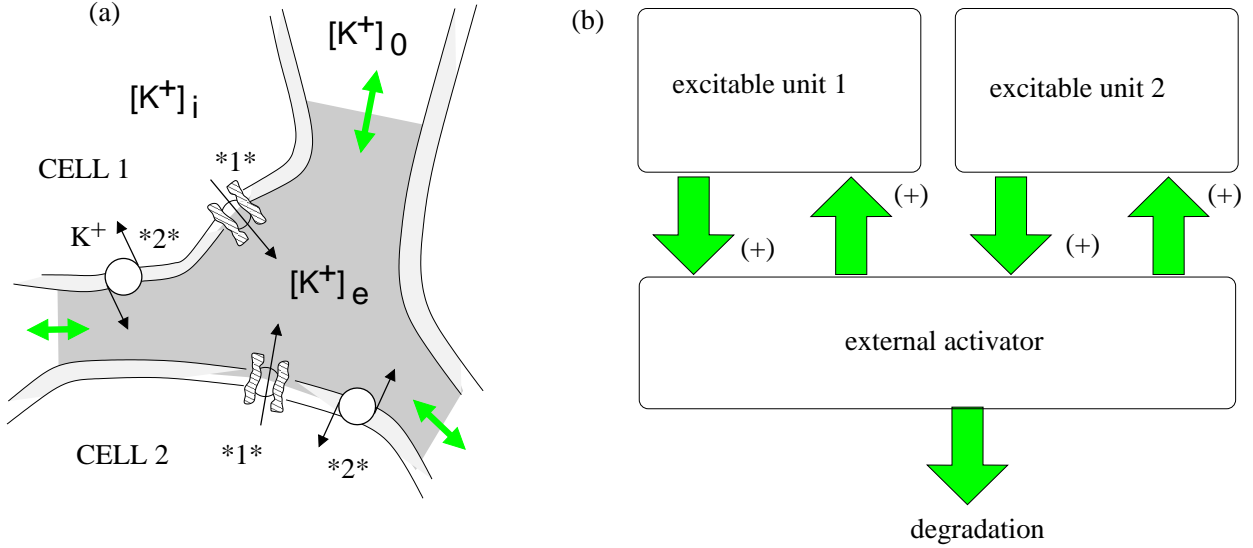}
  \caption{(Color online) (a) Schematic representation of the potassium signaling
    pathways between closely located cells, (b) the corresponding
    structure of the functional model.\label{fig1}}
\end{figure}

We consider an environment which is schematically depicted in
Fig.~\ref{fig1}~(a). We assume that there is a certain volume between
the cells from which the ionic exchange with the outer bath is rate
limited.  For simplicity we assume that this volume is homogeneous and
we denote the potassium concentration here as $[K^+]_e$.

With time, particularly during firing events in neurons, the potassium
channels *1* in its membranes become open and outward currents from
the cells deliver potassium to the extracellular space.  Thus,
$[K^+]_e$ rises while the inter-cellular potassium concentration
$[K^+]_i$ decreases just slightly, because $[K^+]_i \gg [K^+]_e$.  In
fact, one can neglect the associated intra-cellular changes of the
potassium concentration and assume that this concentration remains
constant.

$Na-K$ ATP *2* pumps $K^+$ back into the cells.  This uptake is
balanced by the leakage when the potassium concentration is at
equilibrium value $[K^+]_0$.  The exchange of $K^+$ ions with a
surrounding bath is assumed to take place by a diffusion process,
hence governed by the concentration difference between the exterior
and the bath.  Assuming the bath potassium concentration to be equal
$[K^+]_0$, one can simplify the description of the process by
incorporating all potassium uptake processes in an effective diffusion
rate parameter $\gamma$.

Then the balance of potassium concentration in the inter-cellular space
$[K^+]_e$ can be described as following:
\begin{eqnarray}
\label{e7}
W \frac{d[K^+]_e}{dt}= \frac{1}{ F}\sum_{i=1}^N I_{i,K} + \gamma ([K^+]_0- [K^+]_e),
\end{eqnarray}
where $W$ is the extracellular volume per unit area of the membrane,
$N$ the total number of cells being neighbors to this volume,
$I_{i,K}$ is the electric potassium current per unit area from the
$i$th-cell which is divided by Faraday's constant $F$ to provide the
ion flow.  The second term $\gamma ([K^+]_0- [K^+]_e)$ describes the
effective diffusion of potassium to and from the bath.  This balance
equation (\ref{e7}) provides the basis for the qualitative description
in terms of a functional model we will introduce below.

Note, the variation of the ratio between the extra- and intra-cellular
potassium concentrations affects the corresponding (Nernst-) potential
and, hence, the firing activity.
The considerable rise of extracellular potassium concentration
depolarizes the cell and can evoke the transition to spontaneous
firing.  However, too high extracellular potassium
concentration becomes toxic and can block the cell activity,
completely.

\subsection{Model}
In this subsection we propose a functional model that aims the
qualitative reproduction of main effects arising if a variable
extracellular potassium concentration is taken into account.  The
structure of the model is schematically depicted in
Fig.~\ref{fig1}~(b).  Namely, several excitable units standing for a number
of neurons contribute to the extracellular potassium increase.  This
``external activator'' stands for the inter-cellular space between the
cells with variable potassium concentration.  It is (i) activated
during a high-level state of one of the excitable units, (ii) provides
an additional stimulus to the excitable units and (iii) relaxes to an
equilibrium when receives no activation.

Let us first confine to a single neuron interacting with the external
medium. Particularly, we will implement the activity of the excitable
neuron by the FitzHugh-Nagumo (FHN) system
\cite{FHN,keener}. Therefore we assume that the gating of potassium
release of a single neuron is given by
\begin{eqnarray}
\label{f1}
 \varepsilon \dot{x} &=& x-x^3/3 -y ,\\
\label{f2}
 \tau(x) \dot{y} &=& x + a - C z,
\end{eqnarray}
where $\varepsilon$ controls the time scale separation of the fast
activator $x$ and the slow inhibitor $y$. The operating regime of the
FHN neuron is defined by $a$, playing the role similar to the applied
current in ionic currents-based neuron models.  We assume that it may
fluctuate around some mean value $a_0$:
\begin{equation}
\label{f2a}
a= a_0 + \sqrt{2 D}\xi(t),
\end{equation}
where $\xi(t)$ is white Gaussian noise with zero mean and intensity
is scaled by $D$.

An additional time scale $\tau(x)$ in (\ref{f2}) is introduced in
order to control the two time scales, associated with
firing (high level of $x$ variable) and refractory state (low level of
$x$) independently, which will gain importance in our problem. Specifically, we
introduce the sigmoidal function
\begin{equation}
\label{f4}
 \Psi(x)= \frac{1}{2}\left(1+{\rm tanh}\left(\frac{x}{x_s}\right)\right),
\end{equation}
which is sensitive to the current value of the $x$ variable: it tends
to zero for $x\ll 0$, and to one if $x\gg 0$, while $x_s$ scales the
transition between these states.  Actually, we use (\ref{f4}) as a
smooth replacement of a Heaviside function to distinguish between the
excited and the resting states of the FHN neuron.  With Eq. (\ref{f4})
$\tau(x)$ shapes as
\begin{equation}
\label{f5}
  \tau(x)= \tau_l + (\tau_r - \tau_l)\Psi(x).
\end{equation}
and takes values $\tau_l$ and $\tau_r$ in the rest and excited states,
respectively.

We label the variable dimensionless extracellular potassium
concentration by $z(t) \ge 0$. Its dynamics is given in accordance with
Eq. (\ref{e7}) by
\begin{eqnarray}
\label{f3}
  \dot{z} &=& \alpha g(x) - \beta z.
\end{eqnarray}
with $z=z^0=0$ corresponding to the steady concentration $[K^+]_0$ and
$\beta$ is the rate of ion losses. Respectively the parameters
$\alpha$ stands for the summary ionic fluxes outward the cells. It
scales inversely as well to the size of the extracellular volume.
These fluxes are released into the exterior for high $x$ values if the
cell is excited and channels are open ($x \approx 2$). They disappear
in the rest state $x \approx -1$, if channels are closed. Hence
likewise for the time scales we are able to use the function
Eq. (\ref{f4}) as switcher in Eq. (\ref{f3}). Hence we will set $g(x) =
\Psi(x)$.

The value of $z$ enters in Eq. (\ref{f2}) with a factor $C$. It represents
the depolarizing effect of the increased extracellular concentration.
For a given non vanishing value of $z$ it results in an additional
shift of the $y$-nullcline decreasing effectively the excitability
value $a$. Mathematically we can call it a second activator which was
previously introduced in models for nonlinear semiconductors
\cite{Radehaus_90} .

The set of equations described above is dimensionless and, therefore,
the relationship to ionic current-based neuronal models can be only
qualitative. However, for the sake of simplicity and to keep the
connection with the original problem, we will use the
terminology from neurophysiology further on in order to describe the dynamic
behavior of the model as well as the meaning of control parameters. In
the following we refer to system (\ref{f1})-(\ref{f3}) as the FHN-K
model.

For the numerical simulations the following values of control parameter
values were used: $ \varepsilon = 0.04$, $a_0=1.04$, $C=0.0 \ldots
0.1$ , $\alpha=1.0 \ldots 12.0$, $\beta= 0.05 \ldots 0.5$, $x_0=0.0$,
$x_s=0.2$, $\tau_l=1.0 \ldots 10.0$, $\tau_r=1.0$. It sets the neuron
without the regulation by external potassium ($C=0$) in an excitable
regime.

\subsection{ Nullclines and fixed points of the single unit model}

\begin{widetext}
		
	\begin{figure}[htb]\centering

	\includegraphics[width=\textwidth]{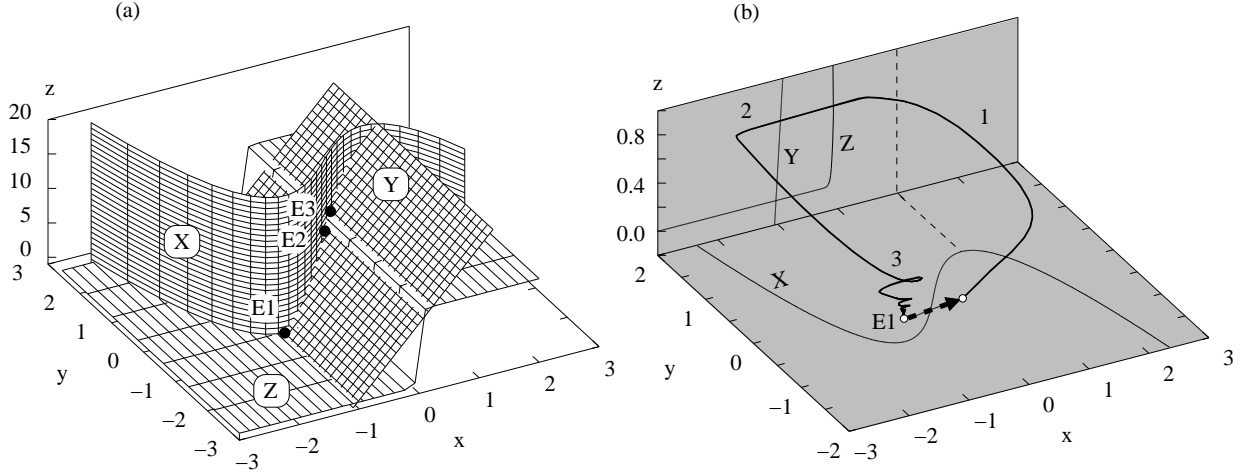}
	\caption{ (a) The 3D plot of nullclines for the FHN-K model.
          The intersections of the cubic $x$-nullcline, linear
          $y$-nullcline and sigmoidal-shaped $z$-nullcline may provide 3
          equilibrium points.  (b) shows a representative trajectory near
          the stable equilibrium point $E1$ (thick line).  The arrow
          indicates the initial perturbation. }
	\label{nulls}
	\end{figure}
\end{widetext}

Let us look for the main features of the model in terms of the steady
states and their stability.  Note, at $C=0$ our model converges back
to the original FitzHugh-Nagumo model with cubic and linear nullclines
and one single equilibrium point, which is stable for $|a_0|>1$ and
unstable otherwise.  Including the $z$-dynamics the model 
(\ref{f1})-(\ref{f3}) possesses three
nullcline surfaces, which are depicted in Fig.~\ref{nulls}~(a) and
labeled $X$, $Y$, and $Z$ according to the equation they satisfy.
One can see, that more than one intersection is possible.  Namely, the
condition $\dot{x}=\dot{y}=\dot{z}=0$ gives for the steady state
values $x^0$:
\begin{equation}
\label{eqpoint}
x^0 +  a_0 =\frac{1}{2}\frac{ C \alpha}{  \beta} \left( 1+{\rm tanh}(\frac{x^0}{x_s}) \right).
\end{equation}
We use a small $x_s$ and the right hand side of (\ref{eqpoint}) becomes 
nearly a step function.  Then one fixed point can be found $x^0_1 \approx
-a_0$, which is always stable for the parameter range we will
consider. Thus the system needs over-threshold stimuli to enhance
states with states $z>0$. 

Taking $C$ as the control parameter regulating the coupling strength
to the exterior variable $z$ two additional fixed points bifurcates at
a critical value $C_\text{crit}$ resulting from
\begin{align}
 	\label{ccrit}
	        \eta C_{\text{crit}} \left( 1+
			\sqrt{1-\frac{2 x_s}{\eta C_{\text{crit}}}} \right) -
			x_s \text{atanh}\left( \sqrt{1-
			\frac{2 x_s}{\eta C_{\text{crit}}}}\right) - a = 0
\end{align}
with $\eta = \frac{\alpha}{2 \beta}$.

\begin{figure}[htb]\centering
	\includegraphics[width=.55 \textwidth]{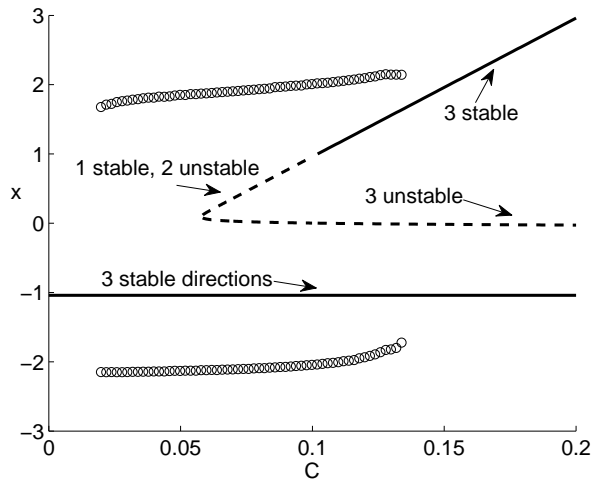}
	\caption{bifurcation diagram for $C$ as the control parameter. 
			The $x$ value of the fixed points is plotted (dashed and solid lines).
			Circles correspond to the extremal elongation $x$-values of
			the limit cycle.			
                        \label{bif}}
\end{figure}

Sections of Eq.~\ref{ccrit} are depicted in Fig.~\ref{bif}. The upper fixed point follows  $x^0_3 = C\alpha / \beta - a_0$ 
and the corresponding $z$-value is the highest $z$-level the system can reach 
($z^0_3  = z_\text{max}= \alpha / \beta$). The fixed point in between is located at 
$x_2 \approx x_s(2\alpha \beta) / (\alpha C - 2\beta xs)$ which is only weakly dependent 
on $C$ and is unstable in every direction.
In  Fig.~\ref{nulls}~(a) they are labeled with $E1$, $E2$, and $E3$,
respectively.

In Fig.~\ref{nulls}~(b) a representative trajectory starting in the
$E1$ vicinity is shown.  The nullcline surfaces are given as
projections on the horizontal and vertical planes.  The arrow
indicates the initial perturbation that kicks the phase point from
$E1$.  After a super-threshold push, the trajectory quickly moves
rightward, then slowly moves along the $y$-direction, but during this
time it also moves upward approaching the $z$-nullcline.  The vertical
component of movement is defined by $\alpha$.  This segment is
labeled with 1.  Within the segment 2 of the trajectory, the vertical
component changes its direction, now it moves downward controlled by
$\beta$.  When the phase point comes back to the vicinity of the
equilibrium point, it is still raised along the $z$ axis. From this
level, the phase point moves downward with pronounced damped
oscillations.  Besides the fixed points there is also a parameter
range where a stable limit cycle appears (see the circles in Fig.~\ref{bif} 
which indicate the extremal elongation in $x$ for the
stable periodic orbit). Initially started outside of the basin of
attraction of the stable fixed points every trajectory ends up on that
limit cycle leading to an oscillatory behavior in $x$ and $y$ and a
nearly constant level of $z$.  Due to a saddle-node bifurcation the
limit cycle looses stability in one direction for too small and too
high values of $C$.  The thereby created saddle-like limit cycle
annihilates with the upper fixed point $E3$ via a Hopf-bifurcation,
illustrated in Fig.~\ref{bif} at the transition from the dashed to the
solid line.

\begin{figure}[t]
  \begin{center}
    \includegraphics[width=.65\textwidth]{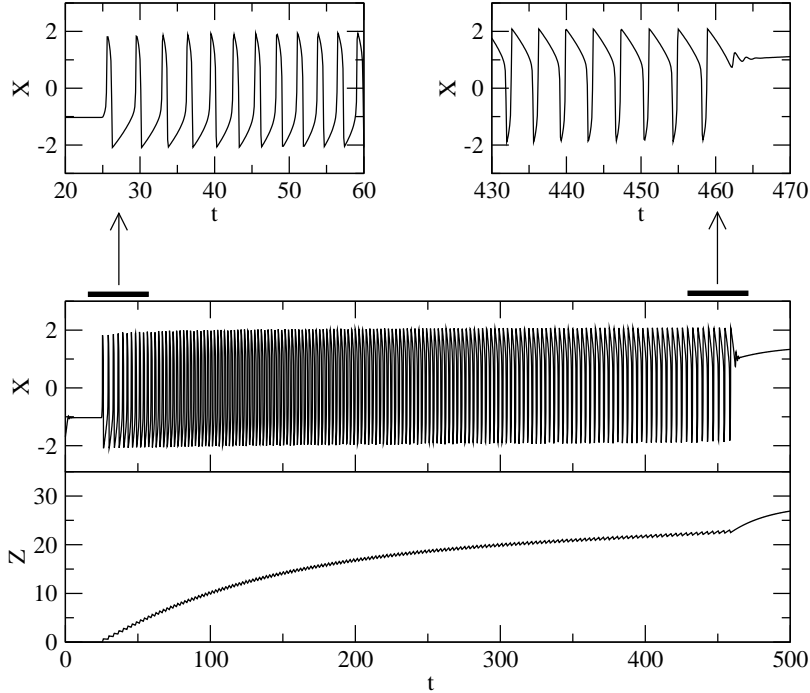}
    \caption{Transition to the stable state at the activation state.
		After initiating the oscillation of $x$, the value of 
		$z$ rises and drives the system 
		to the upper fixed point. In inserts (top row) 
		the corresponding change of the spikes
		for $x$ is illustrated.
		(parameters: $\tau_l=\tau_r=1$, $\beta=0.0354 \ll \alpha=1.0$) 
		\label{antispike}}
  \end{center}
\end{figure}

Beyond the limit cycle in a parameter range, where only the two stable
fixed points exist, $x$ and $y$ also starts to oscillate as a long
living transient when pertubating the lower fixed point initiation as
depicted in Fig.~\ref{antispike}.  In that case the $z$-level
increases successively, lifting the system up to the maximal value.
Thus the depolarizing spikes transform to polarization spikes,
elongating from the depolarized state down to the former resting
polarized level. For the selected set of parameters the level of $z$
still increases. The inverse firing process stops and reaches the
stable steady state $E3$ where neurons are embedded in extracellular
space contaminated by potassium.

\section{ Noisy behavior of a single unit}
\begin{figure}[t] { \centering
		\includegraphics[width=.7\textwidth]{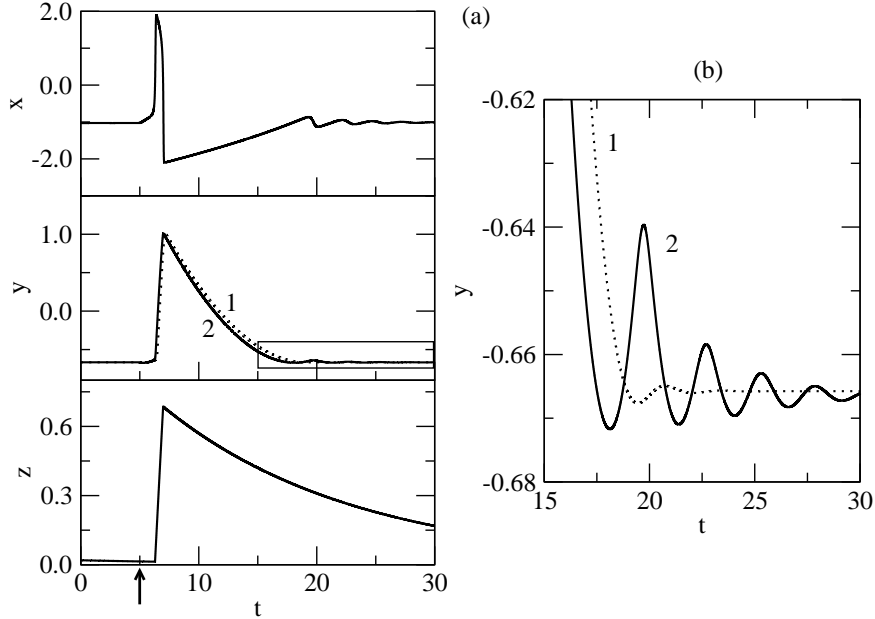}}
		\caption{ (a): Temporal evolution after perturbation in the single-unit model at
		$C=0.008$.  The spike is initiated by a short excitatory pulse
		at the time indicated by the black arrow.  Dotted line (1) in the
		middle panel shows the $y$-time course for $C=0.0$.  (b): the
		enlargement of rectangular area in (a) shows the subthreshold
		oscillations after spiking. The dotted (1) and solid (2) lines
		illustrate the cases of $C=0.0$ and $C=0.008$, respectively.
		\label{fig3} }
	\end{figure}

To understand the specific features of the FHN-K model in a noisy
regime, let us first consider the segment 3 of the trajectory from
Fig.~\ref{nulls}~(b) in terms of the excitation threshold.
Figure \ref{fig3}~(a) shows the time courses of the model variables at
$a=1.004$ when a short external pulse initiates the generation of a
single spike. For specific values of $\alpha$, the activation of
$z$-variable is relatively fast. In the middle panel of Fig.~\ref{fig3}
one can compare the inhibitor behavior for $C=0.008$ (solid curve)
compared with the $C=0$ case (dashed curve). Generally, the solid
curve runs lower after the spike was generated. However, there is a
more complex response near the resting state.

Figure~\ref{fig3}~(b) shows details. As we discussed above, an
increased value of $z$ evokes damped oscillations.  During maxima of
these oscillations the distance to values in phase space where a new
excitation starts is minimal. Hence, during the moments of maximal
elongation a weaker external forcing is sufficient to excite the next
spike.  This feature is important for the understanding how noisy
input acts in this model.  Namely, after a spike was produced and the
refractory time is over, there are few moments of larger probability
for the next noise induced firing.  It resembles the behavior of so
called resonate-and-fire neurons \cite{Izhikevich_01,Tania_04} with
subthreshold oscillations.  But, differently, it occurs only after a
spiking event, if still the exterior with $z \ne 0$ influences the
behavior of the neuron. 

The noisy behavior of model (\ref{f1})-(\ref{f5}) is significantly
controlled by the mechanism described above.  For appropriate small
$D$ values a larger number of noise-induced spikes appears at the first
elongation after approaching the rest state as shown at $t \approx 20$
in Fig.~\ref{fig3}~(b).  There is a kind of stochastic positive
feedback: once appeared, the noise-induced spiking can continue
showing high regularity. 
If at this state current noise values were
to small to override the minimized threshold, then the next spike will
occurs after considerably longer time interval and noisy bursting is
observed \cite{Tania_04,Lacasta_02,schwalger}.

\begin{figure}[t]{\centering
    \includegraphics[width=\textwidth]{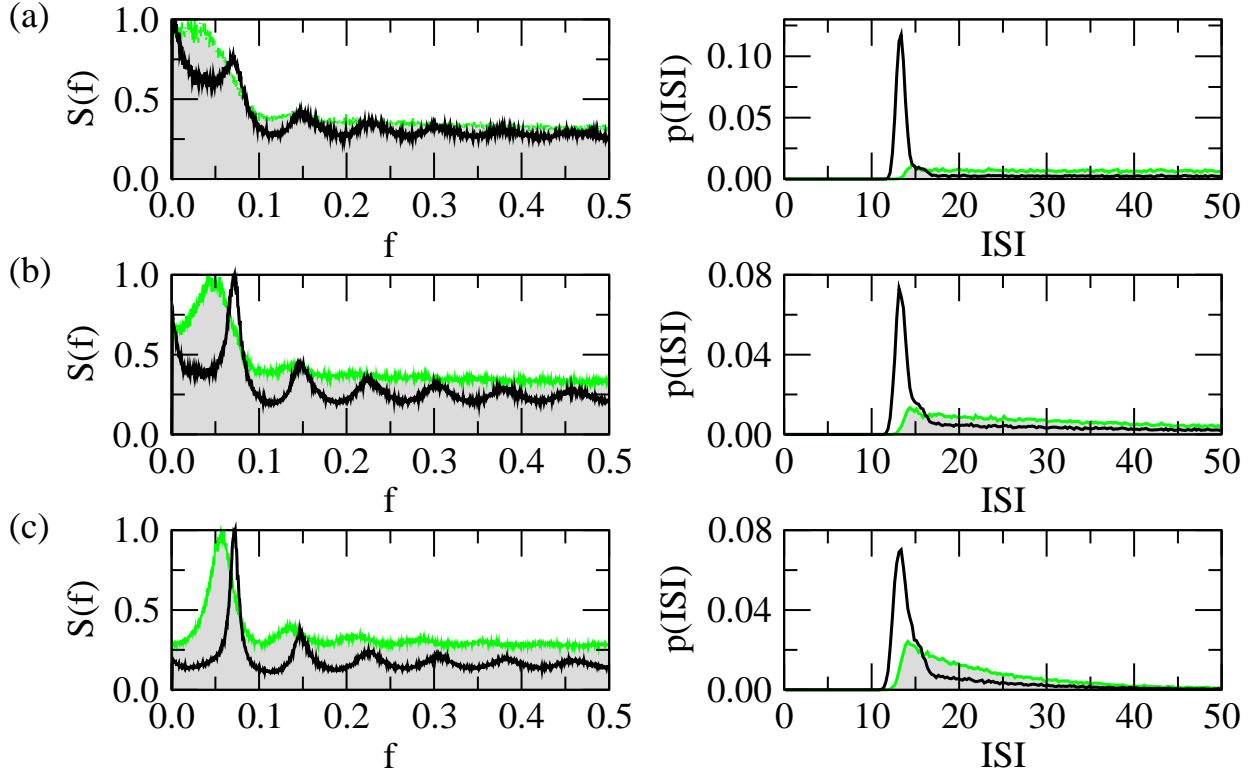}}
   \caption{(Color online) The spectral power density (left panels) and the
     probability distribution density of interspike intervals (right
     panels) for the single-unit model with noise.  For comparison the
     maxima of the spectra are set to one.  Curves in black were
     obtained with $C$=0.03. Curves in green with filled area were
     obtained with $C$=0.0 illustrating the behavior of the
     unperturbed FHN model.  Noise intensity takes values $D$=0.005,
     $D$=0.007 and $D$=0.01 from top to bottom.
     \label{fig4}}
\end{figure}

The described subthreshold oscillations are responsible for specific
features of the averaged spectral power density $S(f)$. In the left
column of Fig.~\ref{fig4} we compare the densities of the unperturbed
FHN model (case $C=0$, given in grey) and of our model at $C=0.03$
(\ref{f1})-(\ref{f5}).  For all panels of the figure, the noise
intensity $D$ is assumed to be small, so the noisy forcing can be
regarded as weak.  Both cases starts with just the same shape of
$S(f)$, when only few spikes appear during the observation time (not
shown in figure).  In the FHN model, the further increasing of $D$
leads to the formation of a broad peak at zero frequency that moves
rightward and reaches the position at $f\approx 0.06$ at $D=0.01$. It
corresponds to the quite regular firing due to the effect of coherence
resonance \cite{review_lindner,pikovsky_1997,han_99}.  Model
(\ref{f1})-(\ref{f5}) shows a similar power spectrum at very weak
and at the final ($D=0.01$) noise strength, while the evolution of the
spectra with increasing noise is different.  Instead of a single broad peak
at zero, two sharp peaks appear at zero and at $f\approx 0.075$.
The inspection of time courses shows, that the first peak corresponds
to the randomly appeared bursts, while the second peak corresponds to
the mean interspike distance within bursts.  With increasing $D$, the
peak at zero gradually disappears, while the peak at $f\approx 0.075$
collects more power.  The third row of panels in Fig.~\ref{fig4} shows
the considerably higher regularity of firing process in the
(\ref{f1})-(\ref{f5}) model comparing with the FHN model. A similar
effect can be observed by inspecting the probability distribution
density of interspike intervals (ISI) as shown in the right column of
Fig.~\ref{fig4}. While the activity near zero frequency is mapped on
interval values with larger than $20$ time units, a pronounced peak
is observed at ISI values $\approx 1$ which grows up to an optimal
value with increasing intensity of noise $D$.

To summarize, the noisy behavior of a single excitable unit is characterized by coherence resonance and by excitation of bursting with (statistically) large intervals between groups of spike. This interesting dynamics is reflected by both the spectral power density and the ISI distribution density. The origin of this feature is what one can call a subsequent self-induced depolarization in interaction with the exterior potassium. Note,  this result is consistent with previously reported behavior of higher-dimensional quantitative models 
\cite{Kager_2000, Kager_2002, Frohlich_2006}

\section{Two excitable units interacting with a common exterior}
The described above self-depolarization plays an important role when
two excitable cells share one reservoir with density $z(t)$. Therefore the equation (\ref{f3}) is replaced by:
 \begin{eqnarray}
\label{f33}
  \dot{z} = \alpha \left( \Psi(x_1) + \Psi(x_2) \right) - \beta z,
\end{eqnarray}
where $x_1$ and $x_2$ belongs to the first and second unit each
described by equations like (\ref{f1})-(\ref{f2}) with the common
variable $z(t)$.

\begin{figure}[t]
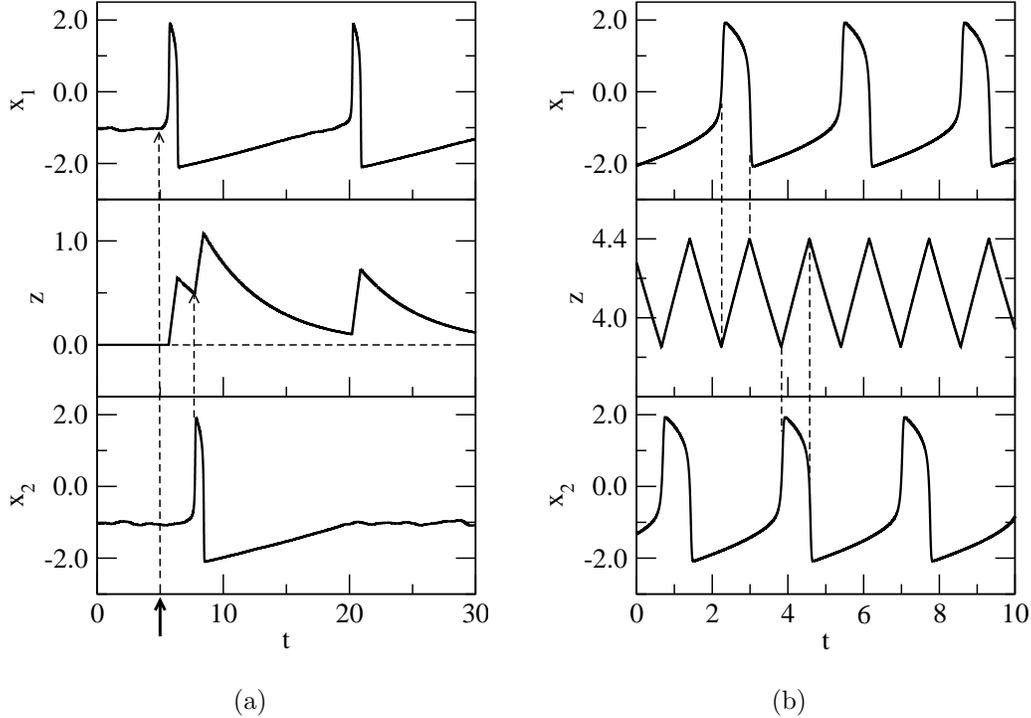

    \subfigure[ ]{\includegraphics[width=0.39\textwidth]{fig7a.eps}}
\hspace{.5cm}
    \subfigure[ ]{\includegraphics[width=0.39\textwidth]{fig7b.eps}}

  \caption{Characteristic operating regimes of two coupled
    excitable units. Left panel: temporal evolution at weak noise.  The first unit with $x_1$
    starts firing first and provides an increase of the common
    external $z$ level.  It depolarizes the second unit and support
    its noise-induced firing.
    Right panel:  Both units are in a oscillatory regime and show firing in anti-phase.
    The frequency of the common output has doubled. The limit cycle of
    these oscillations is in coexistence with the stable fixed point
    $E_1$. \label{fig5}}
\end{figure}

In such a case, firing of one unit provides depolarization for both.
Figure~\ref{fig5}~(a) illustrates the interaction. Originally,
taking separately both units are excitable, so that without external
input they remain in the rest position.  When a noisy stimulus or an
external perturbation initiates a spike in the first unit as shown in
the Fig.~\ref{fig5}~(a), almost instantaneously the second unit
becomes strongly depolarized and starts also firing with large
probability. Considering the collective response, it creates a
doublet, like it is shown in Fig.~\ref{fig5}~(a) at $t\in(5,10)$.  Note,
that the specific time interval in which the second spike appears
depends on the current values of the noisy stimulus.

If the feedback controlled by $C$ is strong enough, a self-sustained
continuous firing occurs after one of the units was excited. In this
regime in case of identical units equally spaced time intervals
between spikes of the first and the second units are adjusted
(Fig.~\ref{fig5}~(b)).  The two units fire perfectly in anti-phase a
effect which is known from glycolytic oscillations in cells
\cite{Heinrich_97}.

Looking forward to the firing patterns in the $2D$ array of many units
discussed in the next section, one can expect both: A temporal shifted
firing of neighboring neurons ("one-induced-by-another" pattern) as
well as a tendency to anti-phase firing of neighboring units. The
latter shows a doubled frequency in the collective response.

\section{ Noisy dynamics of spatially extended models}
Coming to the extended scenario we consider an inhomogeneous medium with separated active units 
on the one hand and exterior modeled by $z$ on the other hand.
There are two main possibilities to construct large ensembles of coupled units in two dimensions defined by (\ref{f1})-(\ref{f5}).

On the one hand on can assume that the $z(\vec{r},t)$ variable describes a continuous diffusive
medium with spatial coordinates $\vec{r}=(r_1,r_2)$. The excitable units are placed in a second 
layer at locally separated sites coupled by the diffusing field $z(\vec{r},t)$.

On the other hand one can alternatively consider a binary medium, consisting of excitable elements 
embedded in the non-excitable field $z(\vec{r},t)$, which diffuses in the remaining 
space of the two dimensional medium.

The first approach is evidently and leads to an usual reaction-diffusion system with three variables. 
Two variables are locally defined and coupled via the third. One might imagine a two layer 
system with excitable units located inside a gel with low connectivity. 
The interaction inside this first layer can be neglected compared to the diffusive coupling 
of the third species $z(\vec{r},t)$.
For such mixed systems with densely packed excitable particles surrounded by reactive 
emulsions pattern formation has been observed in chemical experiments \cite{Vanag_05, Tinsley_09}.

According to the original model of potassium mediated neuronal activity, 
the space is divided on cells, that are electro-chemical active surrounded of 
extracellular space being the diffusive medium for potassium ions. 
Thus, in the present work we will follow the second approach. However, 
we also considered the two-layer system, which we will compare to the 
binary system giving short remarks at the corresponding places.

\begin{figure}[b]
  \begin{center}
  \includegraphics[width=0.7\textwidth]{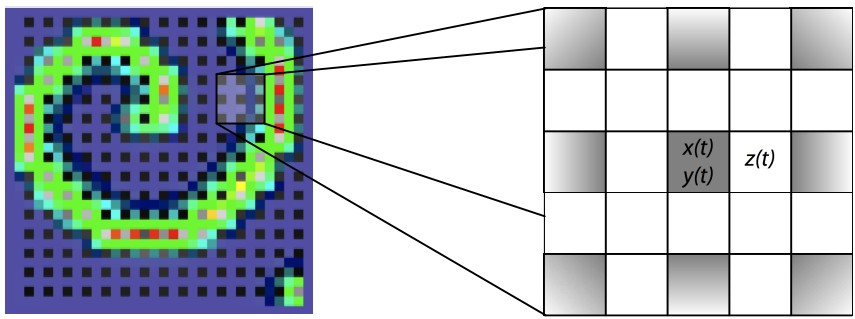}

      \caption{(Color online) Schematic picture of the extended arrangement of neurons ($N$) 
			and pure $z$-cells in between.
		\label{fieldcartoon} }
  \end{center}
\end{figure}

In particular we use a regular  array of active units in each row and column 
illustrated in Fig. \ref{fieldcartoon} following (\ref{f1})-(\ref{f2}). 
For each point of the intermediate diffusive medium it holds:
 \begin{eqnarray}
\label{f333}
  \dot{z}_{ij} = \alpha \sum_{k \in nb_1}\Psi(x_k) +
\gamma \sum_{l,m \in nb_2}(z_{lm}-z_{ij}) -  \beta z_{ij}, \nonumber \\
\end{eqnarray}
where the subscript $ij$ denotes the current point in space.
The second additional term describes the diffusion of the $z$-field with the 
diffusion coefficient $\gamma$. The sum indices $nb_{1,2}$ 
denote the sets of neighboring "neurons" and coupled units, respectively. 
We have implemented several types of coupling like nearest or next-nearest 
neighbor coupling for $nb_1$ and/or $nb_2$. To keep it clear we discuss the 
results only for the coupling to the next $8$ surrounding boxes, where the diagonal elements are 
scaled by a factor of $1/\sqrt{2}$. Taking more neighboring cells into account 
has no mentionable impact. Note, that in contrast to conventional 
reaction-diffusion systems, the active units do not interact mutually 
but only via the common variable $z$, whereas the exterior medium is locally coupled 
with itself and is additionally affected by the neurons activity. 

One can expect that the interplay between the refractory time of an
active cell (controlled by $\tau_l$) and the time scale of
extracellular space (defined by the parameters $\alpha$, $\gamma$ and
$\beta$ in the equation for $z$) plays a key role in the
spatio-temporal dynamics. Therefore, we have selected some
representative examples that are discussed below as five trials.

All the computations are performed in conditions 
where the identical individual units possess at least one stable fixed point 
at $x^0=-a$, which we choose as the initial condition.
The following parameters have been appointed in all cases:
$\varepsilon=0.04$, $a_0=1.04$, $C=0.1$, $\tau_r=1.0$,  $x_0=0$,  $x_s=0.05$. 
Essentially only the parameters of the exterior have been set to different values according to the following table:\\

\begin{center}

\begin{tabular}{|c|c|c|c|c|c|}
  \hline
  Parameters & $\alpha$  & $\beta$    &  $\gamma$  & $\tau_l$  &  $D$ \\
  \hline
  Set \#1   &  50.0  &   6.0  &   2.0  &   1.5  &    0.00005 \\
  \hline
  Set \#2   &  60.0  &  6.0  &   130  &   1.5  &   0.02 \\
  \hline
  Set \#3   &  6.0   &   0.35 &   4.0  &   1.5  &    0.0001 \\
  \hline
  Set \#4   &  10.0  &   0.6  &   0.2  &   2.0  &    0.00002 \\
  \hline
  Set \#5   &150.0  &   6.3  &   2.0  &   1.5  &    0.003 \\
  \hline
  \hline
\end{tabular}
\end{center}
~\\

With the increasing set number the mean $z$-level rises successively
corresponding to a growing value of the coupling parameter $C$.
According to the bifurcation diagram Fig.~\ref{bif} we go to the right
along the abscissa and encounter excitable, oscillatory and bistable
behavior. We underline that the rest state of the uncoupled FHN is
always a stable homogeneous state of our dynamics with $z=0$.
Excitations of spatio-temporal structures need over-threshold noisy
stimuli or corresponding initial conditions.

\begin{figure}[h!]
  \begin{center}
    \subfigure[ ]{\includegraphics[width=0.4\textwidth]{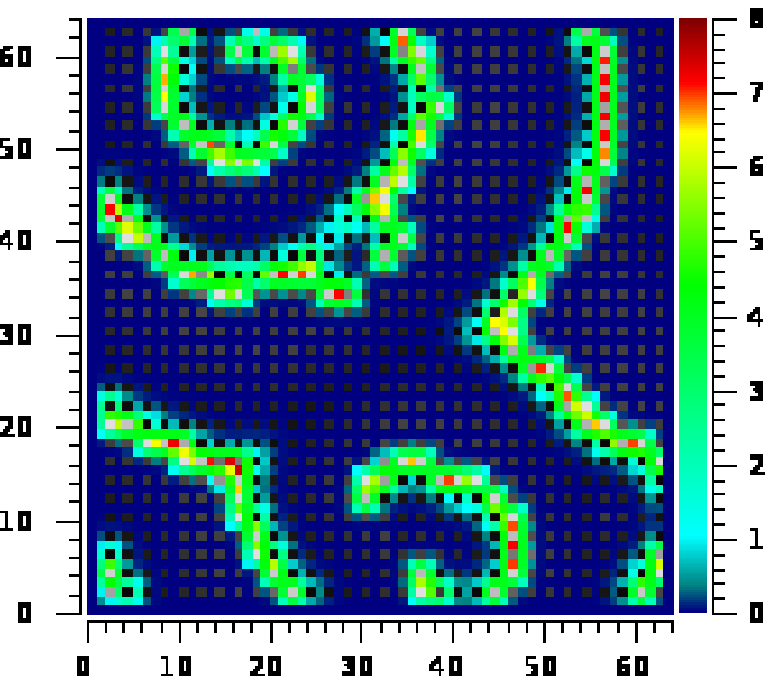}}
    \subfigure[ ]{\includegraphics[width=0.45\textwidth]{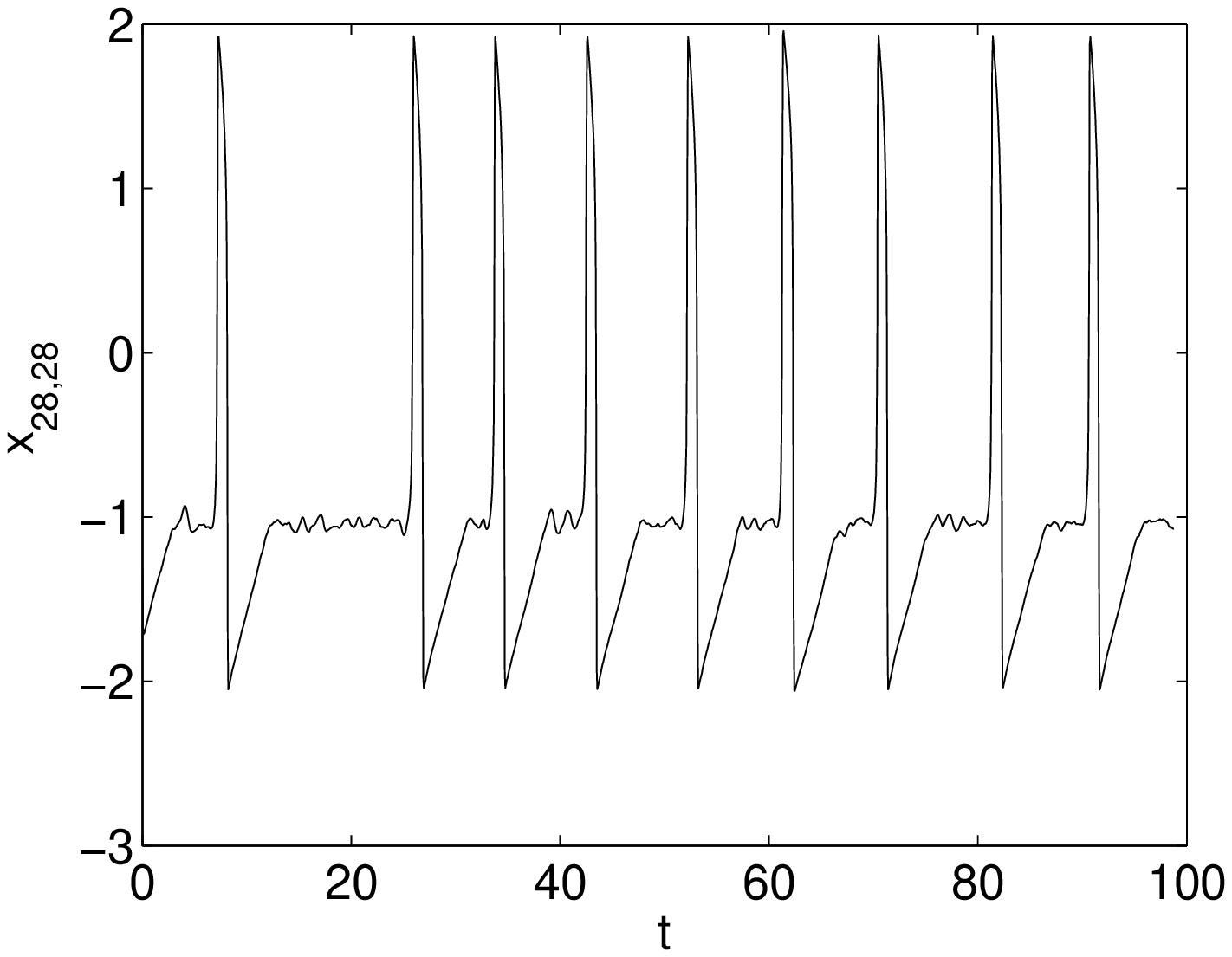}}
      \caption{(Color online) parameter set \#1: (a) Noise induced spirals and 
		wave fronts. The colorbar indicates the $z$-level whereas 
		white and black represents active cells in the 
		excited and rest state state, respectively.
		(b) Time series of an arbitrary chosen cell.
		\label{set1} }
  \end{center}
\end{figure}

\subsection*{Set \#1. Waves, Spots and Spirals
  (Fig.~\ref{set1})}

The local dynamics is excitable. By noise the units can be activated 
and release $z$ to their neighborhood. In this case the outside concentration 
of the medium decays much faster than the units recover, while the diffusion 
is to slow to distribute the delivered $z$ over a large distance.

The units that crossed over in the refractory period ignite other
neighboring units via the medium and a traveling noise-supported
extended waves is excited as it is depicted in Fig.~\ref{set1}~(a).

At the system borders or due to noise circular waves can break and 
the free endings curl to form a spiral wave.
Although the dynamics is purely excitable at the chosen noise intensity 
waves appear very regularly, noticeable in the time series of Fig.~\ref{set1}~(b).
The mean firing rate of the active cells is $r_{mean} \approx 0.1$ and 
the mean exterior concentration is $z_{mean} \approx .8$.

The combination of the chosen diffusion $\gamma=2.0$ with
considerable refractory time of neurons stabilizes the location of the
center of noise-induced spiral wave in spite of all units receive
random stimulus of the same intensity. One may eliminate in the
dynamics the $z$-variable which would reduce the system of equations
to two components standing for a single activator $x$ and inhibitor
$y$. Compared to the usual case the coupling term (diffusion of
activator) appears in the equation of the inhibitor, similar to a
Soret effect. Slightly increased noise strength destroys the spiral
wave structure by splitting it into short fragmented traveling waves
that nucleate and annihilate in a random manner.

The release of potassium in the exterior is still sufficiently low
and so far no new fixed point bifurcates at high $z$-values. Therefore
the background in the large refractory state relaxes always to small
$z$- values as indicated by the dominating blue color in the figure.

In the two-layer system for a decay rate of $\beta \approx 4$ or smaller 
only short living wave segments supported by noise appear.

\begin{figure}[th]
  \begin{center}
    \subfigure[ ]{\includegraphics[width=0.4\textwidth]{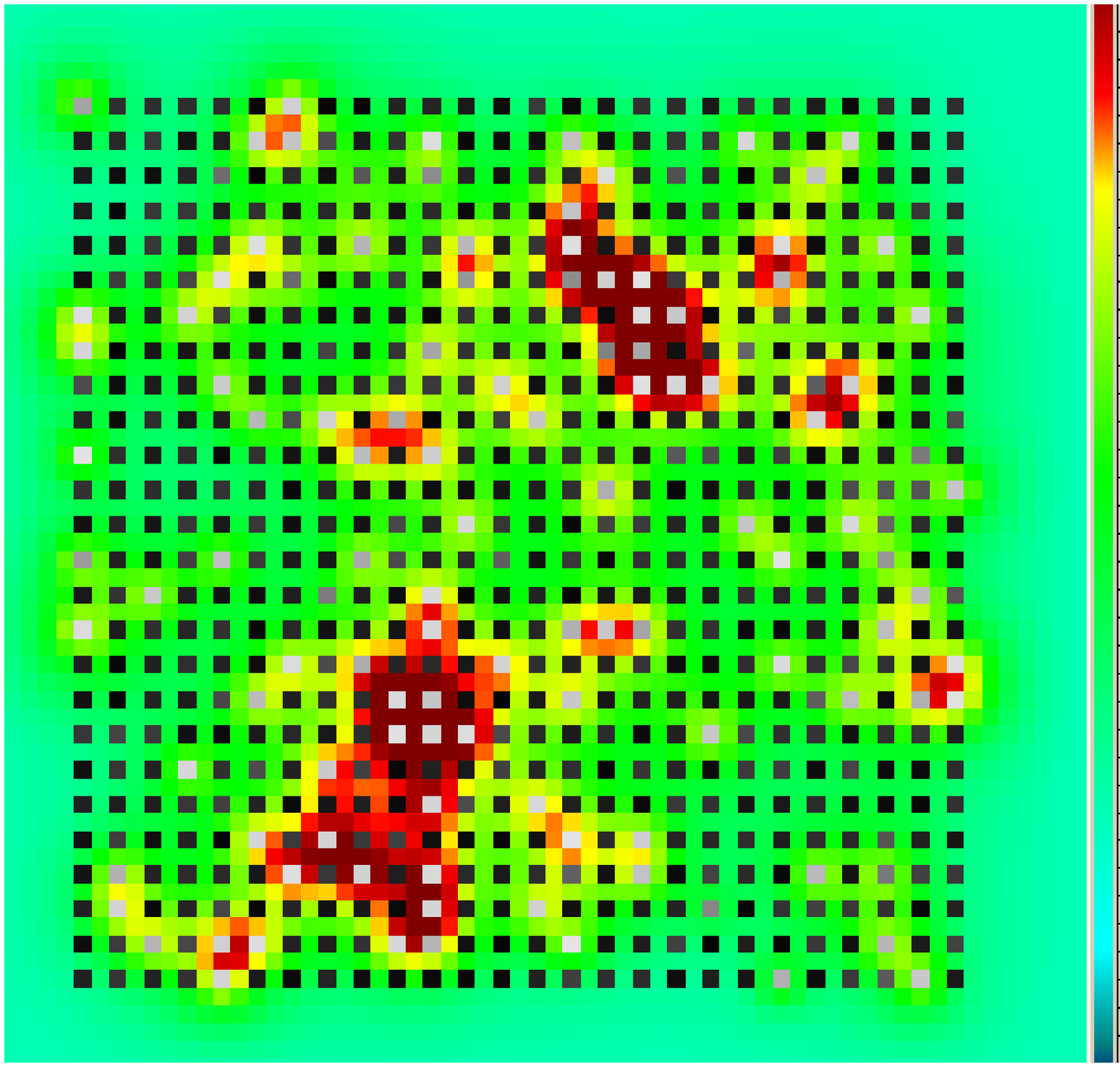}}
\hspace{.5cm}
    \subfigure[ ]{\includegraphics[width=0.4\textwidth]{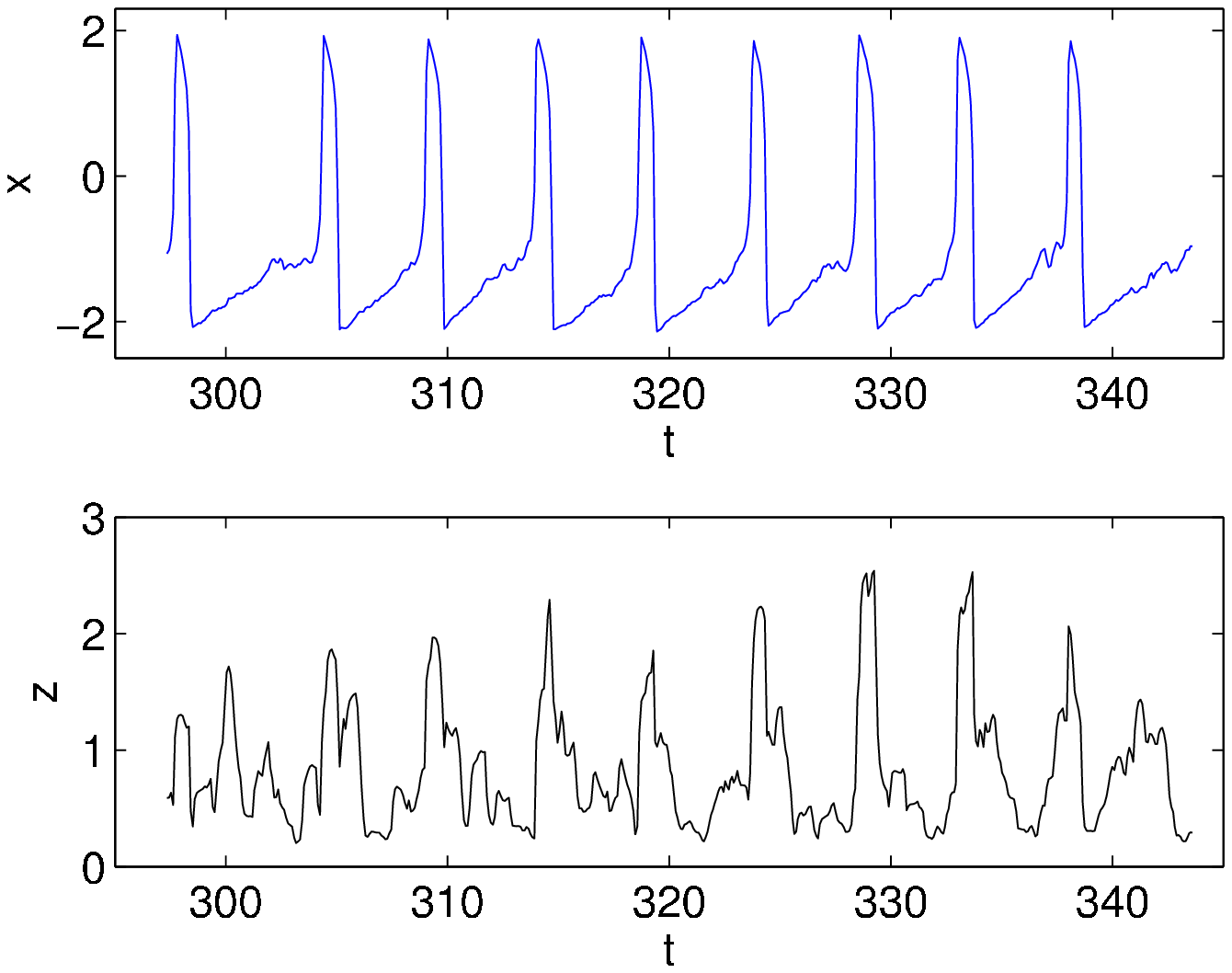}}

    \subfigure[ ]{\includegraphics[width=0.2\textwidth]{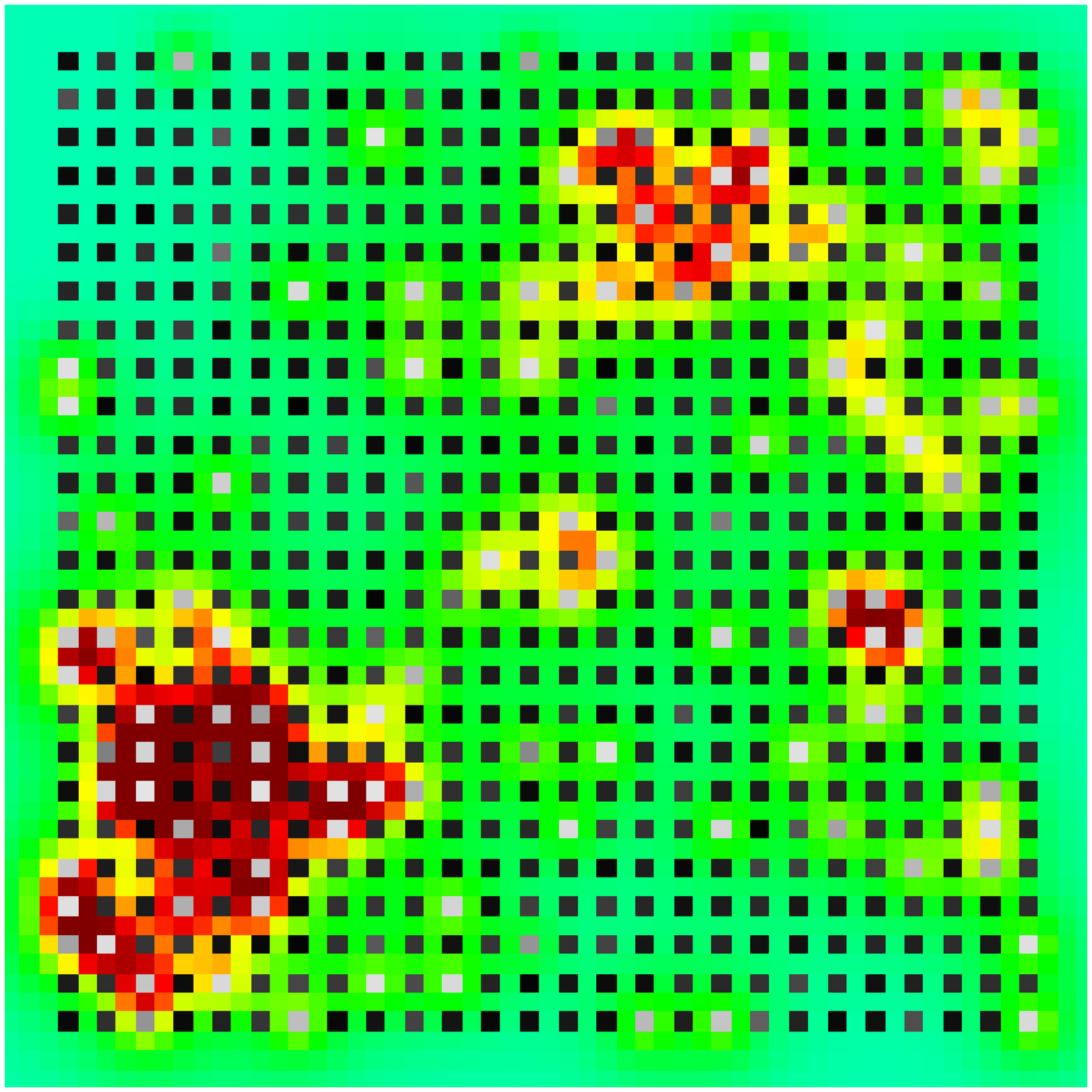}}
    \subfigure[ ]{\includegraphics[width=0.2\textwidth]{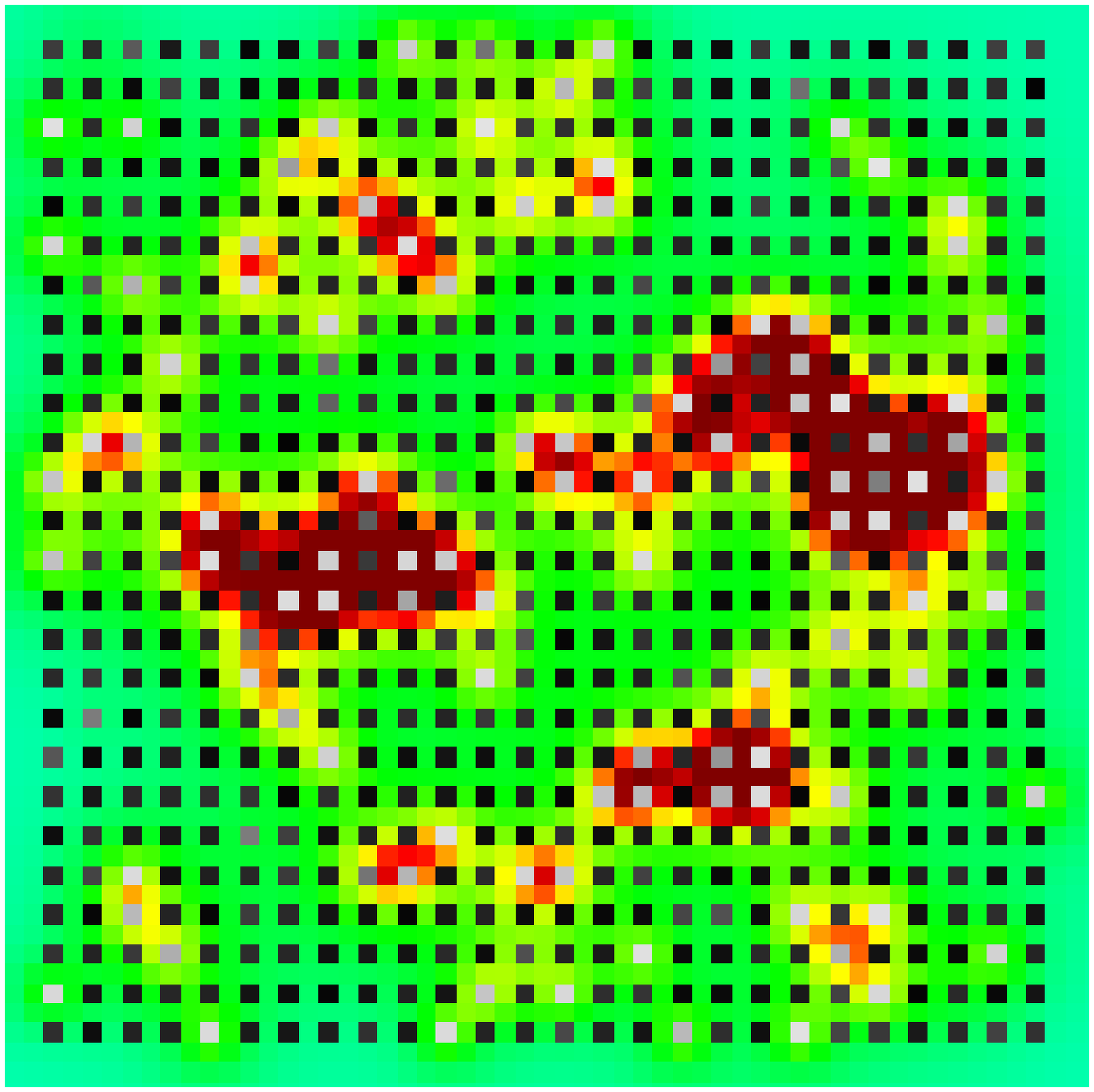}}
    \subfigure[ ]{\includegraphics[width=0.2\textwidth]{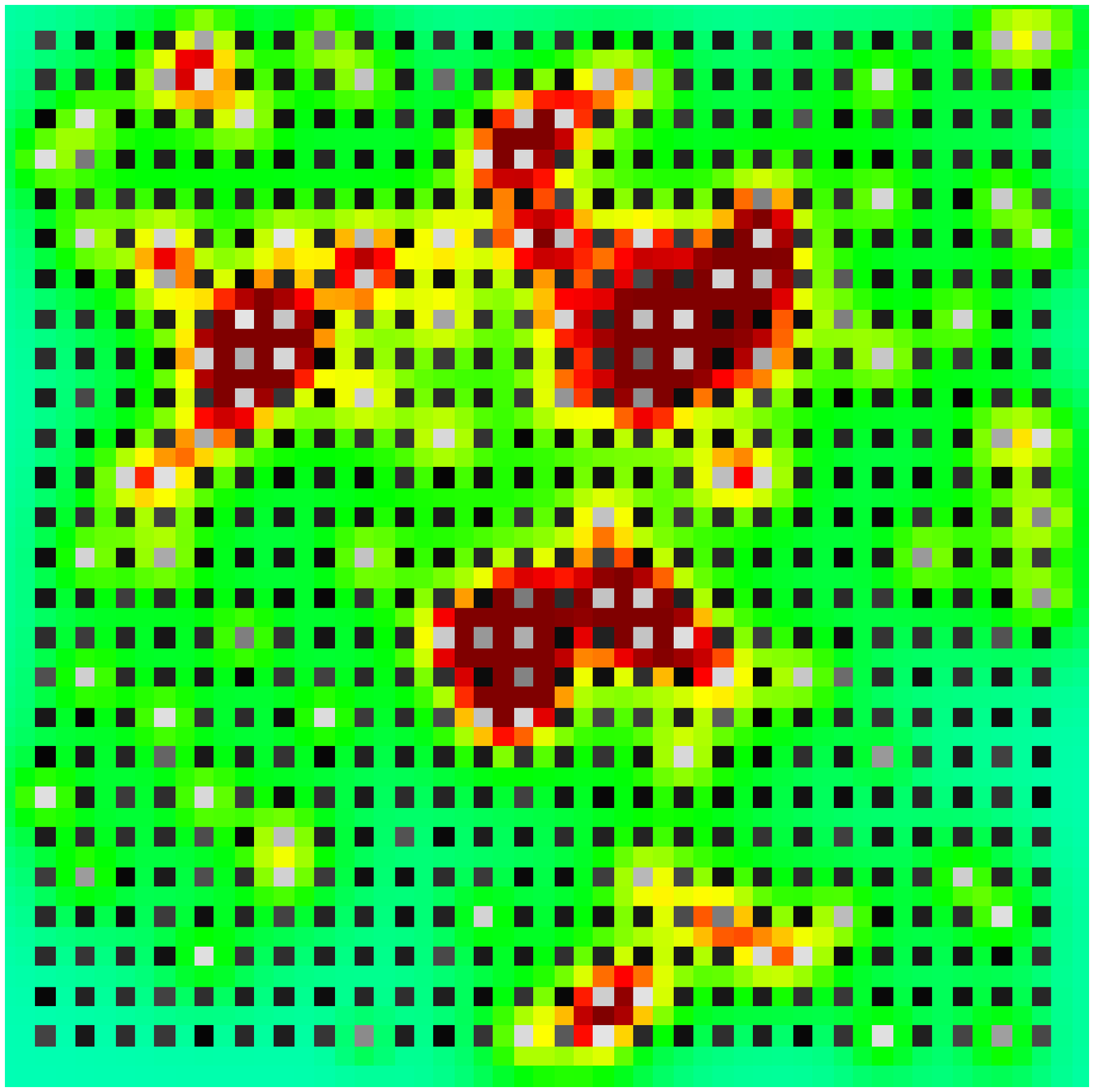}}
    \subfigure[ ]{\includegraphics[width=0.2\textwidth]{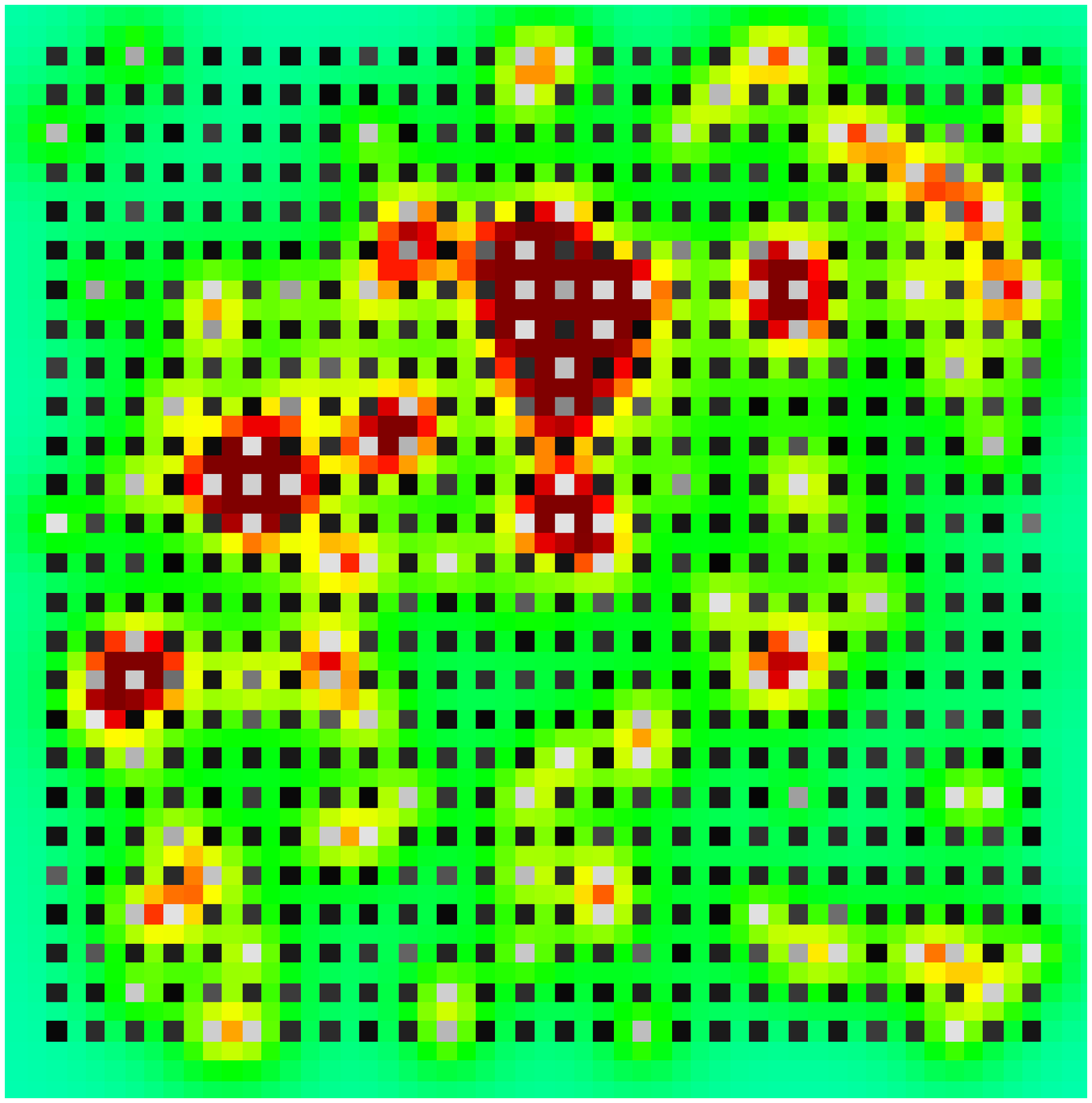}}

      \caption{(Color online) parameter set \#2, 
		 Colors like in Fig.~\ref{set1}~(a)  self-feeding clusters.
		(b) Time series of an arbitrary chosen cell and of an neighboring $z$-cell.
		(c)-(f) snapshots of nucleating, wandering and decaying clusters
		\label{set2}}
  \end{center}
\end{figure}

\subsection*{ Set \#2. Noise supported traveling clusters  (Fig.~\ref{set2})}

The pronounced difference to the former set of parameters is the very
large $z$-diffusion coefficient $\gamma$ and the high noise level.
Nuclei that would lead to coherent patterns like spirals diffuse very
fast forming still connected clusters of delivered $z$. Such developed
clusters can live relatively long wandering through the medium due to
the noisy forcing.

The local dynamics is excitable and possess only the lower fixed point
as a steady state. Due to the fast $z$-diffusion a large vicinity of
units gets activated whenever an unit is excited. It supports the
formation of a localized high-level $z$ region. Inside those areas the threshold
of the units is lowered and the release of $z$ increases. This process
leads to self-feeding meandering cluster as depicted in the snapshots 
of Fig.~\ref{set2}~(c-f).

Note, that noise is necessary over the whole time to keep the clusters alive. Switching the 
noise off lead to a complete decay of the $z$-level to zero. 
For the given noise intensity the stochastically occurring spikes events are quite regular.
The $z$-level follows the activation and forms also coherent oscillation-like elongation 
as shown in Fig.~\ref{set2}~(b).
The mean rate is $r_{mean} \approx 0.2$ and the mean $z$-concentration is $z_{mean} \approx 1.0$.

\begin{figure}[h]
  \begin{center}
    \subfigure[ ]{\includegraphics[width=0.4\textwidth]{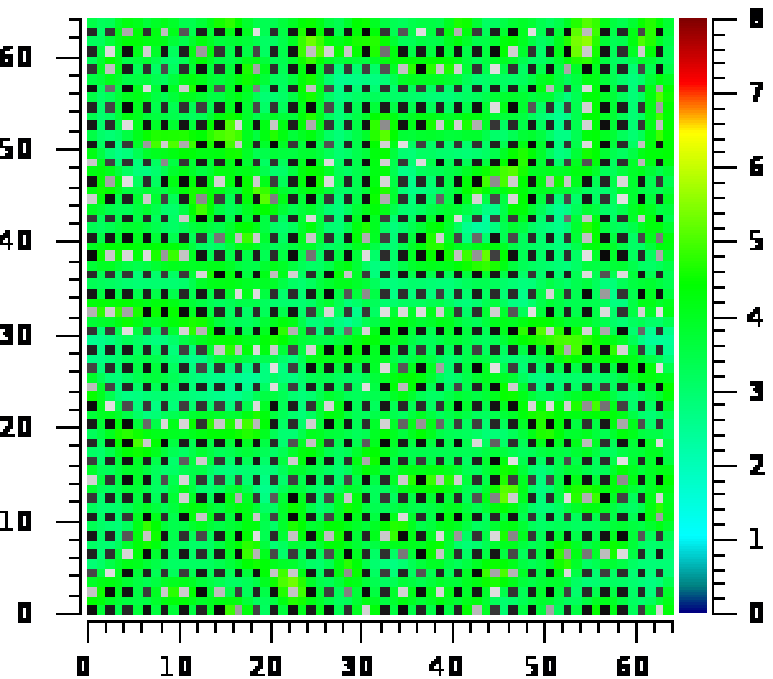}}
    \subfigure[ ]{\includegraphics[width=0.45\textwidth]{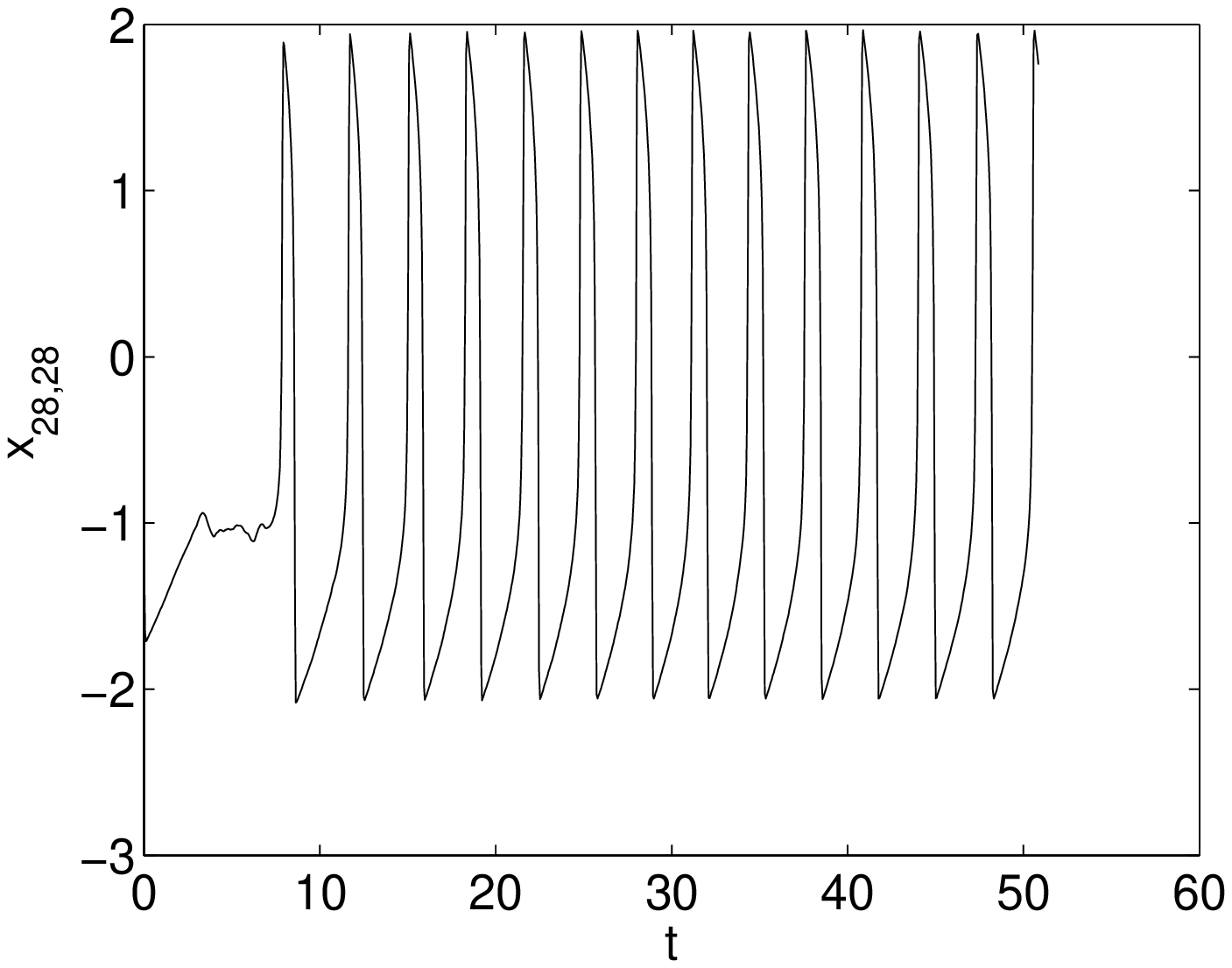}}
      \caption{(Color online) parameter set \#3: (a) Elevated $z$-level 
			due to permanently oscillating active cells.. 
		 Colors like in Fig.~\ref{set1}~(a).
		(b) Time series of an arbitrary chosen cell.
		\label{set3}}
  \end{center}
\end{figure}

\subsection*{ Set \#3. Desynchronized oscillators embedded in a $z$-sea (Fig.~\ref{set3})}

Compared to the former case less potassium is released but it is also
slower decaying. The refractory
time is short $\tau_l = O(1)$ compared to the decay time $\beta^{-1}$
and therefore we observe oscillating units (Fig.~\ref{set3}~(b)) embedded
in a situation that high $z$ values survive longer than the duration
of one oscillation period. Thus the exterior is permanently fed by
potassium which is diffusing over long distances, shown in
Fig.~\ref{set3}~(a).

Starting at the $z=0$ level, the active units first perform 
the noise induced transition to the oscillatory behavior.
Except for the the initiating perturbation noise is not needed 
to keep the oscillation alive. All units moves along the stable limit 
cycle but with different phases.
Along such sites the medium is quickly filled with $z$ which 
starts to propagate elevating the neighborhood and forming a 
front like spread over the space.
It is a typical scenario of nucleation in systems with multiple
attractors. 

Despite that a second attractor at high $z$-values exists it will never be reached due to
the decay rate $\beta$ which is still high enough to compensate 
the release of potassium with the rate $\alpha$. 
Thus a quasi-steady picture remains
with a sea of high potassium populated by active units blinking
regularly and feeding the exterior with potassium. 
Here, we find for the oscillation frequency $r_{mean} \approx 0.34$ and 
for the mean exterior $z_{mean} \approx 5.7$.

Two times greater
$\alpha$ and $\beta$ and $20$ times smaller diffusion 
originates another similar pattern: after some transient, 
high $z$-concentration occupies all
available space. After that, a noise-induced neuronal firing moving
as complex pattern through the medium.  A detailed inspection shows
that the observed behavior is locally based on the anti-phase firing of
neighboring units inside the constant sea of high $z$-level.  In 2D
space it produces a chess-like firing pattern, when all neighbors fire
in anti-phase.  In the deterministic case it forms regular phase waves
moving from borders to the center.  Noise adds irregularity and evokes
many additional patterns.

In figure \ref{korrelation}a the spatial correlation function is shown for 
an arbitrary neuron over the distance to its neighboring excitable units 
along a row. The solid line represents the long range correlation to the 
active units in the neighborhood, when the noise has not yet disturbed 
the wave-like propagation of the stimulus that lifts the units to the limit cycle.

A slow decay of the correlation is shown expressing the 
indirect diffusive coupling. The first dip corresponds to 
the next nearest neuron which is less correlated to the considered 
unit than the next but one. It reflects that on average neighboring 
elements can fire preferable in anti-phase.

However, a small amount of noise will drive the system to a complete 
desynchronized state after a couple of oscillations, shown as the dashed line 
in Fig.~\ref{korrelation}~(a)

The described situation is typical for the studied extended system and can be found 
over a large parameter range. Also in the two-layer system the same oscillating regime 
exists for the same parameter set.

\begin{figure}[h]
  \begin{center}
    \subfigure[ ]{\includegraphics[width=0.4\textwidth]{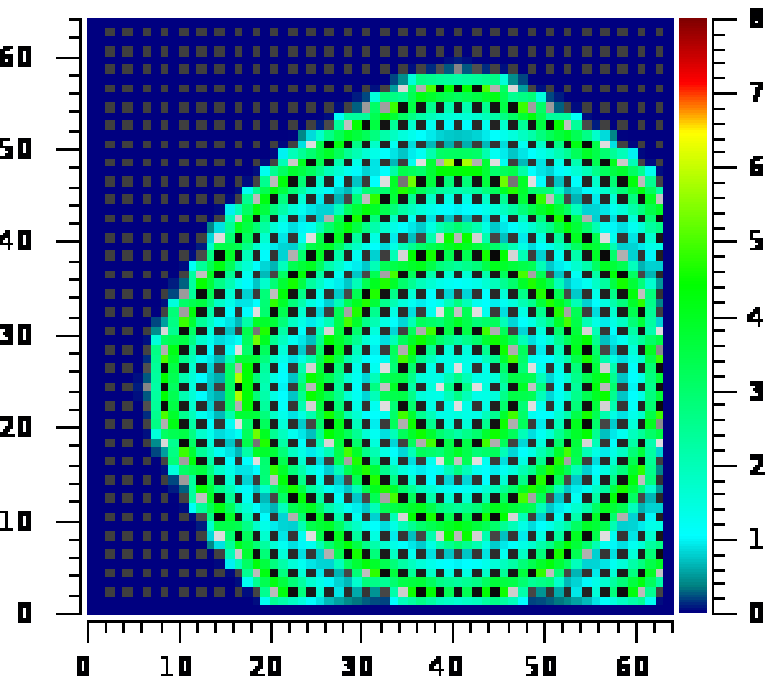}}
    \subfigure[ ]{\includegraphics[width=0.45\textwidth]{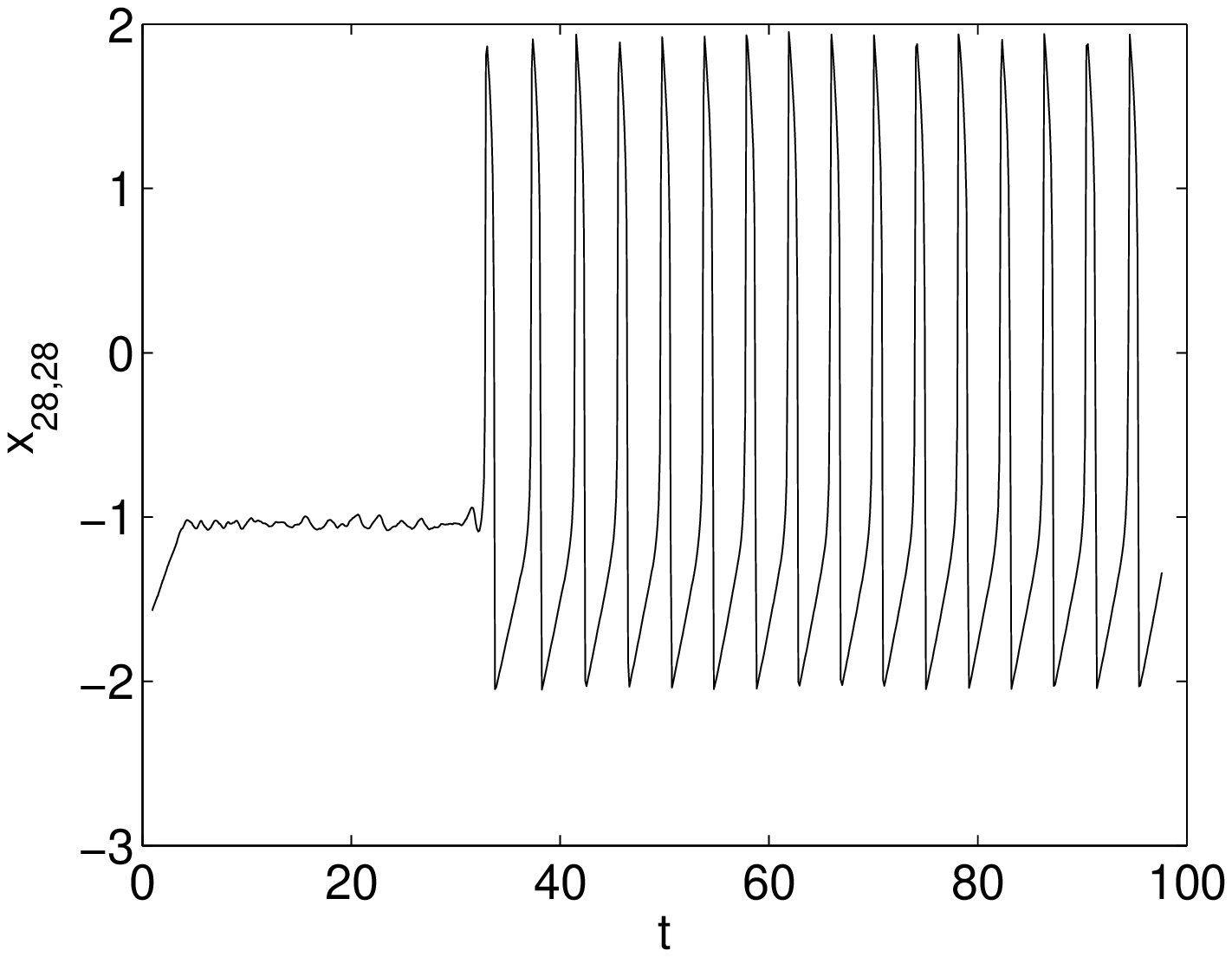}}
      \caption{(Color online) parameter set \#4: (a) Noise induced concentric waves. 
		Colors like in Fig.~\ref{set1}~(a).
		(b) Time series of an arbitrary chosen cell.
		\label{set4} }
  \end{center}
\end{figure}

\subsection*{Set \#4. Oscillations form a propagating ring-like pattern (Fig.~\ref{set4})}

Similar to the former parameter set, a single cell, 
fluctuating around the rest state, can reach the stable
limit cycle, we mentioned above, by overcoming the unstable limit
cycle due to noise. At this noise level those events are rare. Once
happened $\beta$ is so small, that the $z$-level around the
oscillating cell can rise and reach the next cells without decaying
before. Therefore all active units can be elevated to the oscillatory
behavior successively and a concentric wave appears, shown in
Fig.~\ref{set4}~(a).  A typical time series of $x$ recorded from a
single cell is depicted in Fig.~\ref{set4}~(b). For the chosen
parameters the oscillation period after the transition is $r_{mean}
\approx 0.25$, while the $z$-level averages $z_{mean} \approx 6.0$.
Compared to the last case diffusion of potassium is reduced
drastically. It gives reason that spatial structure can establish at
length scale of a few neurons. 

\begin{figure}[h]
	\begin{center}
		\subfigure[ ]{\includegraphics[width=0.45\textwidth]{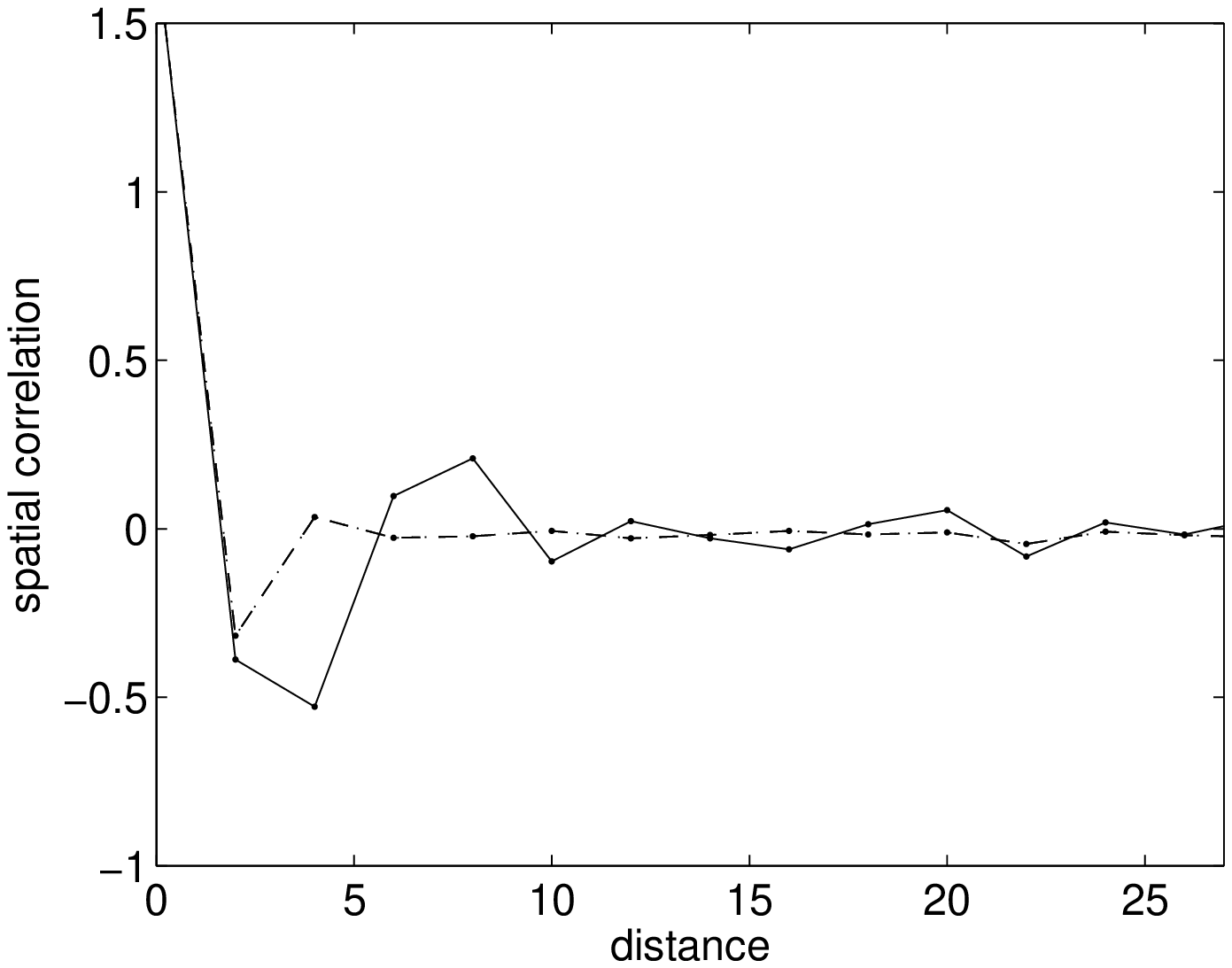}}
		\subfigure[ ]{\includegraphics[width=0.45\textwidth]{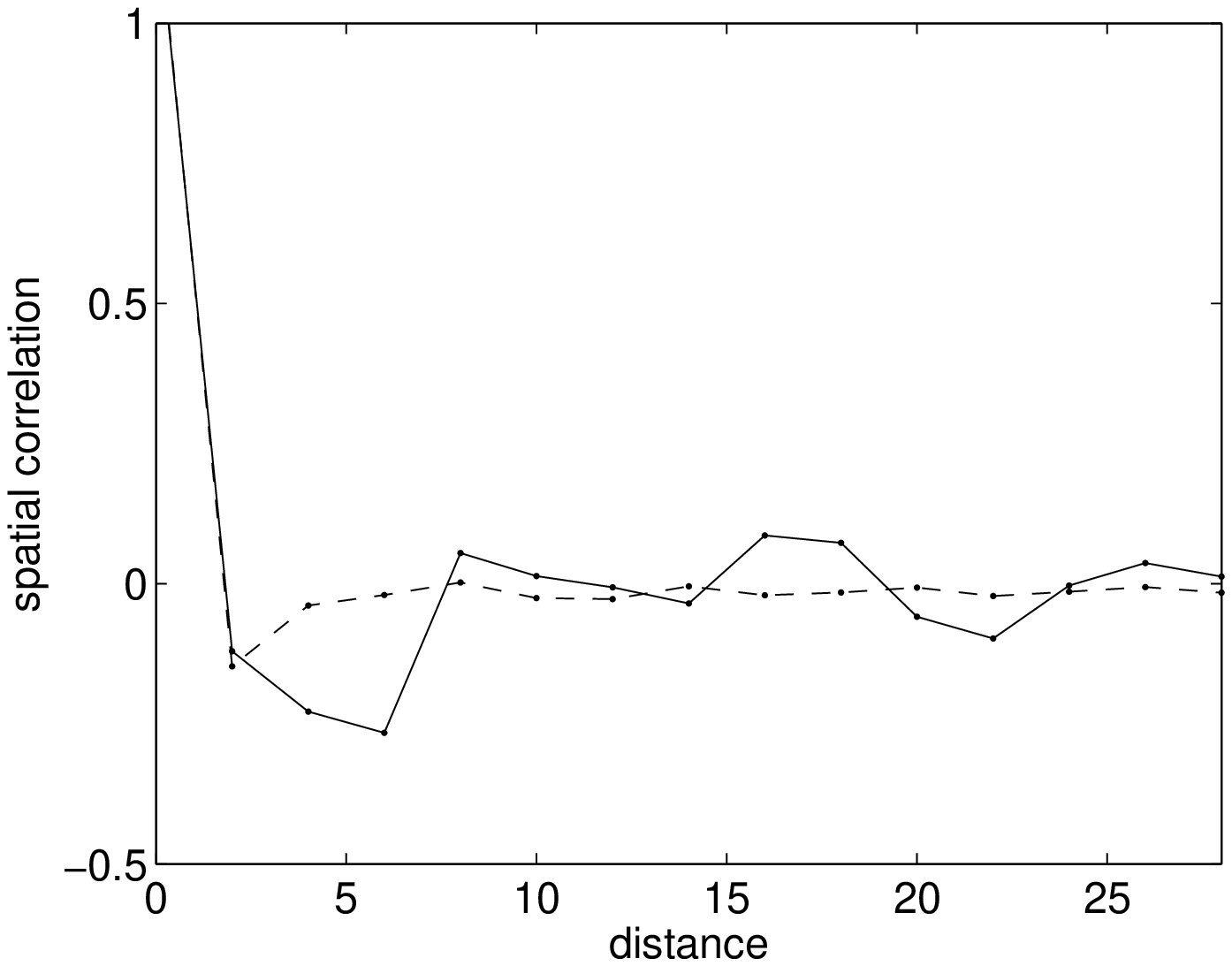}}

		\caption{The spatial correlation of set \#3 (left) and set  \#4 (right) over the 
				distance to neighboring neurons is shown without considering 
				the inactive space. Solid lines 
				indicates transient behavior for early times, dashed lines show 
				the system which becomes uncorrelated due to noise.
			\label{korrelation}}
	\end{center}
\end{figure}

In Fig.~\ref{korrelation}~(b) the spatial correlation is shown, 
where the solid line shows a long range 
correlation shortly after initiating the wave pattern. The active units 
are well synchronized and it takes longer time until the structure 
is destroyed by noise. This state corresponds to the 
dashed line in Fig.~\ref{korrelation}~(b).

Increasing $\beta$ the stable and unstable limit cycle annihilate 
and the local dynamics is excitable. A noise induced super-threshold 
perturbation leads to a singular depolarization of the active cell and 
the $z$-level in it's neighborhood increases. Thus the next cell becomes
depolarized as well and a singular ring-like wave emerges.

Further increase of $\beta$ allows only a small neighborhood 
of the initially activated cell to get enough released $z$.
However, such activated wave segments can be stable a long time 
while traveling through the medium.

This kind of patterns always decays in the two-layer situation.
The released $z$ can diffuse to each site of the array without 
restriction. So the release rate $\alpha$ needs to be greater.
For example choosing $\alpha = 15$ and a big enough initial nucleus 
an oscillon is formed. This structure is an extended but localized spot fed by 
oscillating cells inside and surrounded by inactive cells.

\begin{figure}[t]
  \begin{center}
    \subfigure[ ]{\includegraphics[width=0.4\textwidth]{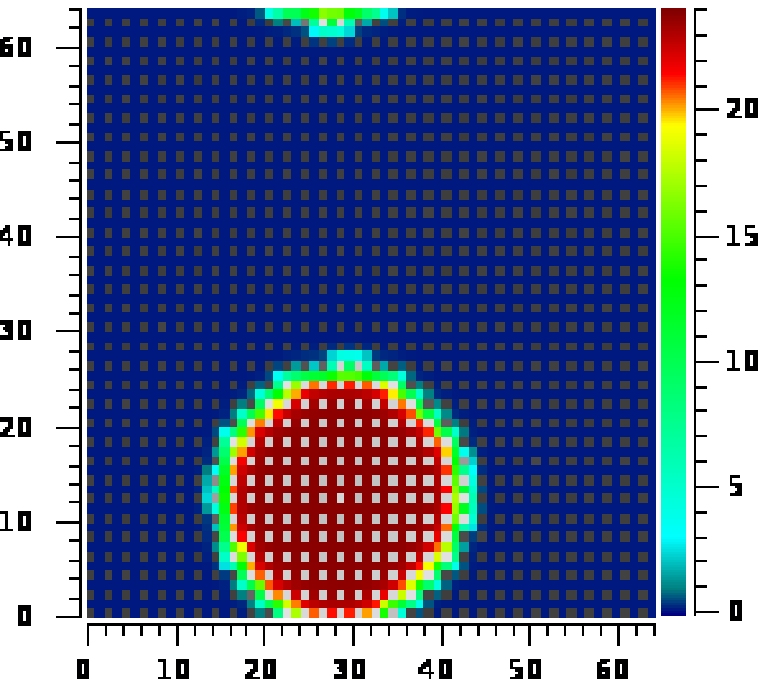}}
    \subfigure[ ]{\includegraphics[width=0.4\textwidth]{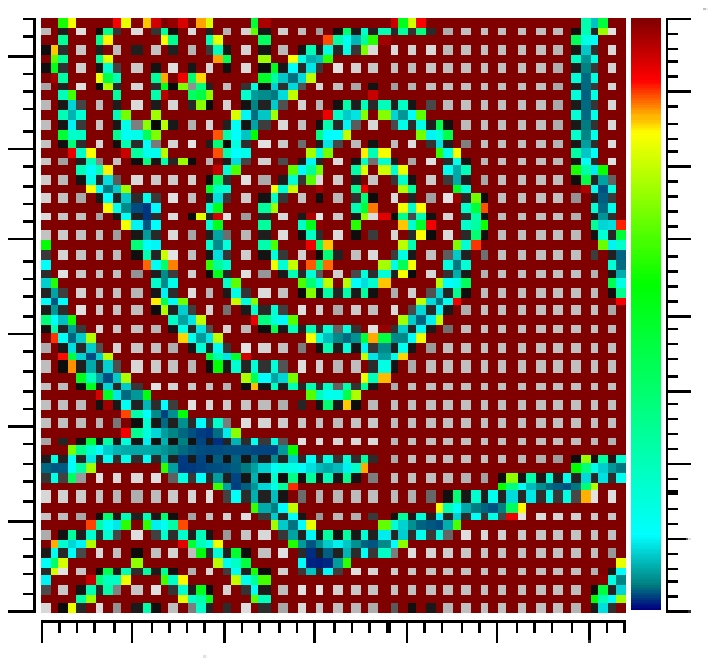}}
      \caption{(Color online) parameter set \#5: (a) A bistable wave 
		covers the medium with the high-level $z$ state. 
		 Colors like in Fig.~\ref{set1}~(a).
		(b) Noise induced inverted spirals and waves with polarized states appear. 
		The scales are the same like in (a).
		\label{set5}}
  \end{center}
\end{figure}

\subsection*{ Set \#5. Bistability and inverted waves. (Fig.~\ref{set5})}

Figure \ref{set5} illustrates the case of the highest release rate
$\alpha$, considered here. 
After the nucleation of a bistable wave 
all the space becomes initially occupied by
the high-level $z$ state ($z=\alpha/\beta \approx 24$) which is stable due to the bifurcation of 
a second stable fixed point at the depolarized state of the cells. 
Thus the vicinity of the activated cells is permanently filled with 
$z$ dissipated with $\beta$ and diffusively distributed with $\gamma$.
An inverse situation occurs. Noise can create low level patterns, like 
inverted spirals or propagating waves shaped 
into the high-level $z$ sea. They consist of polarized states 
which propagate through the medium. Such coherent patterns 
does not exist in the vicinity of the 
considered parameters for the two-layer system, after the 
high-level $z$ state is reached.
Only independent break-ins are taking place stochastically.

Depending on the noise configuration 
it is also possible, that the system reaches the $z=0$ state again 
by the appearance of the inverted bistable front. 

Other types of unconventional patterns have been previously reported. 
These are rotating spirals or target waves which run from outward to the center 
called antispirals or antiwaves, respectively.
Such patterns have been found in the Belousov-Zhabotinsky 
reaction and elsewhere and can be described by reaction-diffusion systems 
\cite{Vanag_01,Shao_09,Baer_04}. 
Note, that in our case the rotational direction of the spirals 
and the propagation of the waves is the same as for 
common waves like presented in Set \#1. 

From the dynamical point of view, the existence of this regime can
be explained as follows. The considerable elevation of $z$ shifts the
operation point of core FHN model to the opposite side of the cubic
nullcline. Operating on the background of high exterior $z$ and
corresponding to the shifted fixed point, each subthreshold
perturbation leads to an inverted spike being the impulse from resting
high level to the low one. Now $\tau_r$ controls the duration of
an polarization spike while $\tau_l$ define the refractory time (Fig.~\ref{set5}~(b))
The above described mechanism is already discussed in Fig.~\ref{antispike} 
for a single active unit having two stable steady states.

\section{Conclusions}
In this paper we have introduced a model that qualitatively describes
the neuronal dynamics at variable extracellular concentration of
potassium ions. Using the FitzHugh-Nagumo model as prototype for an
excitable unit, we added a pathway that qualitatively takes into
account the potassium release from the neurons and the depolarization
(threshold lowering) as a result of the increased extracellular
potassium level.

The analysis of the model for a single unit, for two coupled units, as
well as for an extended array have shown that:

\begin{itemize}
 \item[(i)] The stochastic model of a potassium driven excitable neuron
exhibits bursting as the firing pattern. Its due to subthreshold
oscillations for higher values of potassium. As a result, the power
spectrum with increasing intensity differs from the typical one for an
excitable system with strong time scale separation. It shows an 
pronounced and narrow peak at a frequency different from zero. The
broad peak gradually close to zero-frequency disappears with growing
noise intensity.

 \item[(ii)] In the excitable regime two potassium-coupled neurons being
forced by noise show doublets of spikes. Firing of one neuron strongly
depolarizes the second neuron and makes its firing almost inevitable.
In many cases these doublets tend to an oscillatory behavior being
synchronized in anti-phase and possessing a frequency doubled to the
spiking of a singular neuron. 

 \item[(iii)] A two dimensional array of potassium-driven neurons shows a
variety of noise-induced spatial-temporal firing patterns depending on
the relation between the characteristic timescales of the model and
the noise intensity.  One of the most interesting patterns is the
long-living randomly-walking spots of depolarized states. 
Another effect is the high-level potassium state with inverted 
spirales of the polarized state being the result of the medium-driven shift of
equilibrium state toward the right part of cubic nullcline.
\end{itemize}

In spite of simplicity of the generalized model we use, some links can be made between our results and  relevant  neuro-physiological studies. Namely, the well known but still
 debating 'potassium accumulation hypothesis'  
\cite{Green_1964,Fertziger_1970,Frohlich_2008} considers the  self-sustained  rise of  extracellular potassium as the cause of epileptiform activity. 
Taking it in mind, our computational results can be classified as following:
  the short-term activation of z-medium (including concentric and
 running waves) might describe the potassium dynamics within the  physiological range,
 while the patterns with persistent high  level of $z$ variable resemble the  
formation of epileptic seizure and thus  can be  regarded as representing pathological conditions.
Thus, an interesting future work can be done to reveal  the possibility and conditions for mutual transitions between 'normal' and 'pathological' states. 

We thank S. R\"udiger (Berlin) for fruitful discussion and cooperation.
D.P. acknowledges the support from RFBR grant 09-02-01049.
F.M. and L.S.G. acknowledges the support by the SFB 555 and 
R.B. S. and L.S.G. thanks for support by the Bernstein Center for 
Computational Neuroscience.


\begin{thebibliography}{99}
\bibitem{HH} Hodgkin A.L., Huxley. A.F.,
			 J. Physiol. London, {\bf 117}, 500-544 (1952).

\bibitem{keener} Keener J. and Sneyd J.,
			 {\em Mathematical Physiology}  (Springer, New York) (1998).

\bibitem{FHN} FitzHugh R.A., 
			Biophys.  J. {\bf 1}, 445 (1961).

\bibitem{review_lindner} Lindner B., Garcia-Ojalvo J., Neiman A.B., 
			and  Schimansky-Geier L., Phys. Rep.392, 321 (2004).

\bibitem{sykova_83} Sykova E. 
			Prog. Biophys. Mol. Biol. {\bf 42}, 135 (1983).

\bibitem{Dahlem_03}  Dahlem Y.A., Dahlem M.A., Mair T., Braun K. and M\"uller S.C., 
			Exp. Brain Res. {\bf 152}, 221 (2003)

\bibitem{Yi_2003} Yi C.-S., Fogelson A.L., Keener J.P. and Peskin C.S.
 			{\bf 220}, 83 (2003).

\bibitem{yan_96} Yan G.X., Chen J., Yamada K.A., Kleber A.G. and Corr P.G.
			 J.Physiol. {\bf 490}, 215 (1996).


\bibitem{hansen_78} Hansen A.J. 
			Acta Physiol. Scand. {\bf 102}, 324 (1978).


\bibitem{barreto_1} Cressman J.R. Jr., Ullah G., Ziburkus J., Schiff S.J. and Barreto E.
			J. Comput Neurosci {\bf 26}, 159-170 (2009).


\bibitem{barreto_2} Ullah G., Cressman J.R., Barreto E. and Schiff S.J.
			J. Comput Neurosci {\bf 26}, 171-183 (2009).


\bibitem{Deitmer} Deitmer J.W. , Rose C.R., Munsch T., Schmidt J. ,
			Nett W.  , Schneider N.-P.  and Lohr C. , 
 			Glia {\bf 28}, 175-182 (1999).



\bibitem{Bazhenov1} Bazhenov, M., Timofeev, I., Steriade, M. and
			Sejnowski, T.J.,
			Journal of Neurophysiology {\bf 92}, 1116-32 (2004).

\bibitem{Bazhenov2} Park E-H. and Durand D.M., 
			Journal of Theoretical Biology {\bf 238}, 666-682 (2006).

\bibitem{Vern_1977} Vern B.A., Schuette W.H., Thibault L.E.,
			J Neurophysiol 40(5):1015-23. (1977)

\bibitem{Gardner-Medwin_1983} Gardner-Medwin A.R.
			J Physiol 335:393-426,(1983).

\bibitem{Odette_1988} Odette L.L., Newman E.A.
			Glia 1(3):198-210, (1988).

\bibitem{Dietzel_1989} Dietzel I., Heinemann U., Lux H.D.  
			Glia 2(1):25-44. (1989).

\bibitem{Kager_2000} Kager H., Wadman W.J., Somjen G.G.. 
			 J Neurophysiol 84(1):495-512 (2000).

\bibitem{Kager_2002} Kager H., Wadman W.J., Somjen G.G.. 
			J Neurophysiol 88(5):2700-12, (2002).

\bibitem{Frohlich_2006} Fr\"ohlich F., Bazhenov M., Timofeev I., 
			Steriade M., Sejnowski T.J.  
			J Neurosci 26(23): 6153-62, (2006).

\bibitem{Frohlich_2008} F. Fr\"ohlich, M. Bazhenov, V. Iragui-Madoz
  			and T. J. Sejnowski, 
			Neuroscientist 14 422 (2008).

\bibitem{potass_1}Postnov D.E., Ryazanova L.S., Sosnovtseva O.S.,
			Mosekilde E.,
			Journal of Neural Systems{\bf 16}, 99-109 (2006) .

\bibitem{potass_2}Postnov D.E., Ryazanova L.S.,Zhirin R.A. , Mosekilde E.
			and Sosnovtseva O.V.,  
			International Journal of Neural Systems, Vol. 17, No. 2 (2007)
 			105-113

\bibitem{coombes_03}Timofeeva Y. and Coombes S. , 
			Phys. Rev. E  {\bf 68}, 021915 (2003).

\bibitem{coombes_06}Timofeeva Y., Lord G.J. and Coombes S. , 
			Neurocomputing {\bf 69}, 1058-1061 (2006) 


\bibitem{Epstein}Berenstein I. ,Dolnik M. ,Yang L., Zhabotinsky A.M. and Epstein I. R.: 
			Phys. Rev. E {\bf 70}, 046219 (2004).


\bibitem{Showalter}Taylor A. F. , Tinsley M. R. , Wang F., Huang Z. and Showalter K.:
			Science {\bf 323} 614-617 (2009)

\bibitem{Falcke}Falcke M. 
			2003 New J. Phys. {\bf 5} 96  . 

\bibitem{Jung} Shuai J.W. and Jung P. , 
			Phys. Rev. E  {\bf 67}, 031905 (2003).

\bibitem{Radehaus_90} Radehaus C., Dohmen R., Willebrand H. and
			Niedernostheide F.-J.,
			Phys. Rev. A {\bf 42}, 12 (1990).


\bibitem{Izhikevich_01} Izhikevich E.M.
			Neural Networks J. {\bf 14},883-894 (2001).

\bibitem{Tania_04} Verechtchaguina T., Schimansky-Geier L., and
			Sokolov I. M.,
			Phys. Rev. E {\bf 70}, 031916 (2004).


\bibitem{schwalger}Schwalger T. and Schimansky-Geier L.,
			Phys. Rev. E {\bf 77}, 031914 (2008).

\bibitem{Lacasta_02} Lacasta A. M., Sagu{\`e}s F., and Sancho J. M.,
			Phys. Rev.  E {\bf 66}, 045105 (2002).


\bibitem{pikovsky_1997} Pikovsky A.S., Kurths J., 
			Phys. Rev. Lett. {\bf 78}, 775-778 (1997).


\bibitem{han_99} Han S.K., Yim T.G., Postnov D.E., 
			and Sosnovtseva O.V., 
			Phys. Rev.  Lett. {\bf 83}, 1771 (1999).


\bibitem{Heinrich_97} Wolf J. and Heinrich R., 
			BioSystems {\bf 43}, 1-24 (1997).

\bibitem{Vanag_05} Epstein I. R., Vanag V. K., 
			Chaos {\bf 15}, 047510 (2005).


\bibitem{Tinsley_09}Tinsley M. R., Taylor A. F., Huang Z. and
			Showalter K., 
			Phys. Rev. Lett. {\bf 102} 1583301 (2009).


\bibitem{Green_1964} Green .JD., 
			Physiol. Rev.  44:561-608.(1964).

\bibitem{Fertziger_1970} Fertziger A.P., Ranck J.B. Jr. 
			Exp.  Neurol. 26(3):571- 85 (1970).


\bibitem{Vanag_01} Vanag V. K. and Epstein I. R.,
			Science {\bf 294}, 835-837 (2001).

\bibitem{Shao_09} Shao X., Wu Y., Zhang J., Wang H., and Ouyang Q.,
			Phys. Rev. Lett. {\bf 100}, 198304 (2008).

\bibitem{Baer_04} Nicola E.M., Brusch L., and B\"ar M.,
			J. Phys. Chem. B {\bf 108 (38)}, 14733-14740 (2004)


\end{thebibliography}
\end{document}